\documentclass[12pt]{article}
\usepackage[latin9]{inputenc}
\usepackage{amsmath}
\usepackage{amssymb}
\usepackage{esint}
\usepackage[numbers,square,sort&compress]{natbib}
\usepackage[unicode=true,pdfusetitle,
 bookmarks=true,bookmarksnumbered=false,bookmarksopen=false,
 breaklinks=false,pdfborder={0 0 1},backref=false,colorlinks=false]
 {hyperref}
\hypersetup{
 pdfborderstyle=}

\makeatletter

\DeclareTextSymbolDefault{\textquotedbl}{T1}

\numberwithin{equation}{section}

\usepackage{amsfonts}
\usepackage{ytableau} 
\usepackage[square,numbers,sort&compress]{natbib}

\makeatother

\begin{document}
\begin{center}
\textbf{\large{}Unfolded Dynamics Approach and Quantum Field Theory}{\large\par}
\par\end{center}

\begin{center}
 
\par\end{center}

\begin{center}
\vspace{0.2cm}
 \textbf{Nikita~Misuna}\\
 \vspace{0.5cm}
 \emph{Max-Planck-Institut für Gravitationsphysik (Albert-Einstein-Institut),}\\
 \emph{ Am Mühlenberg 1, 14476, Potsdam, Germany }\\
 \vspace{0.5cm}
 \textit{Tamm Department of Theoretical Physics, Lebedev Physical
Institute,}\\
 \textit{Leninsky prospekt 53, 119991, Moscow, Russia}\\
\par\end{center}

\begin{center}
\vspace{0.6cm}
 nikita.misuna@aei.mpg.de \\
\par\end{center}

\vspace{0.4cm}

\begin{abstract}
\noindent We study quantization of a self-interacting scalar field
within the unfolded dynamics approach. To this end we find and analyze
a classical unfolded system describing $4d$ off-shell scalar field
with a general self-interaction potential. Then we systematically
construct three different but related unfolded formulations of the
corresponding quantum field theory, supporting them with illustrative
calculations: an unfolded functional Schwinger--Dyson system, an
unfolded system for correlation functions and an unfolded effective
system for vertex functions. The most curious feature we reveal is
that an unfolded quantum commutator gets naturally regularized: a
standard delta-function is replaced with a heat kernel, parameterized
by the unfolded proper time. We also identify an auxiliary $5d$ system,
having this proper time as a physical time, which generates $4d$
scalar action as its on-shell action.

\newpage{}

\tableofcontents{} 
\end{abstract}

\section{Introduction}

Quantum field theory (QFT) represents a powerful and elaborated theoretical
framework, that unifies quantum mechanics and special relativity.
However, it is widely believed that solving the puzzle of quantum
gravity will require a radical paradigm shift, since the direct application
of QFT methods to general relativity does not lead to a meaningful
theory.

One possible way out is to add new degrees of freedom while increasing
the symmetry of the theory. A natural attempt is to consider theories
with higher-spin fields. From the point of view of standard QFT, interactions
with such fields seem problematic: if higher-spin (HS) fields are
massive, corresponding interactions are in general non-renormalizable;
if HS fields are massless, then related gauge symmetries turn out
so restrictive that at first sight they forbid any interactions at
all (for a review of HS no-go theorems see \citep{NoGo}). However,
thanks to symmetries, in both cases we have remarkable examples of
profound and highly nontrivial theories. String theory contains infinite
sequences of massive HS fields, but infinite-dimensional superconformal
symmetry organizes them in such a way that the whole theory is UV-finite.
Vasiliev HS gravity describes massless fields of all spins and possesses
a certain infinite-dimensional HS gauge symmetry governed by Fradkin-Vasiliev
algebra \citep{FradVas}, but this symmetry requires a non-zero value
of the cosmological constant, so the theory is formulated in anti-de
Sitter space, where it eludes the taboos of the no-go theorems. For
a partial review of the recent literature related to HS problems see
\citep{snow}.

Available formulations of Vasiliev HS gravity represent generating
systems for classical equations of motion, written within the framework
of the unfolded dynamics approach \citep{unf1,vas1,vas2,unf2,ActionsCharges}.
This includes Vasiliev theories in $4d$ \citep{vas1,vas2}, in $3d$
\citep{hs_3d}, Vasiliev theory for symmetric bosonic fields in arbitrary
dimensions \citep{hs_arb_d}, chiral HS gravity \citep{xir1,xir2}
etc. All this models are purely classical.

One of the main problem of HS gravity is that its nonlinear action
is unknown, although some alternatives for an action principle have
been put forward, see e.g. \citep{act1,act2,act3,act4,act5,act6,act7,act8}.
This provides an obstacle to the systematic study of quantum HS gravity.
Nevertheless, there is a number of results on quantum higher spins
in the literature, in particular, on 1-loop partition functions \citep{Z1,Z2,Z4,Z5,Z6,Z7,Z8,Z9,Z10},
on amplitudes \citep{amp0,amp1,amp2,amp3,amp4,amp6,amp7}, on the
finiteness of chiral HS gravity \citep{chir1,chir2,chir3,chir4,chir5,chir6}
etc. In this paper we address the problem of quantization of the field
theory within Vasiliev's unfolded dynamics approach, which does not
directly use a classical action. Namely, we provide a systematic procedure
for an unfolded quantization of a $4d$ self-interacting scalar field.

To this end we first present a classical unfolded formulation for
this model. This formulation is interesting by itself, because the
number of available unfolded nonlinear theories is very limited by
now. To quantize it, we use the method proposed in \citep{misuna},
where it has been shown that quantization can be performed via identifying
a certain submodule of the off-shell unfolded system with an external
source, conjugate to the unfolded field, and promoting classical unfolded
equations to unfolded Schwinger--Dyson ones (other suitable classical
off-shell unfolded systems, that allow for such identification, include
free integer-spin fields in $4d$ Minkowski \citep{misuna} and anti-de
Sitter \citep{misuna-1} spaces, as well as free chiral and gauge
supermultiplets in $4d$ Minkowski space \citep{misuna3}). For related
proposal of the so-called Lagrange anchor in application to the unfolded
field theory see \citep{Anchor1}. Developing the idea of \citep{misuna},
we construct three unfolded formulations of the corresponding quantum
field theory: unfolded functional Schwinger--Dyson system, an unfolded
system for correlation functions and an unfolded effective system
for vertex functions.

The paper is organized as follows. In Section \ref{SEC_UDA} we give
a brief overview of the unfolded dynamics approach and present and
analyze a classical unfolded system for $4d$ self-interacting scalar
field. In Section \ref{SEC_SD} we quantize this unfolded system by
deducing unfolded functional Schwinger--Dyson equations that determine
a partition function of the theory, and use them to calculate a free
unfolded propagator. In Section \ref{SEC_CORREL} we present a closed
system of equations on unfolded correlation functions and with its
help evaluate a first perturbative correction to the unfolded propagator.
In Section \ref{SEC_LOOP} we formulate a prescription for the system
of unfolded effective equations and consider a particular realization,
calculating a one-loop correction to the inverse unfolded propagator.
In Section \ref{SEC_HOL} we present a toy $5d$ model, which leads
to the classical $4d$ theory under consideration and reveals some
curious quasi-holographical features. In Appendix A we present a general
consistency analysis of the classical unfolded system for the self-interacting
scalar field and comment on differences between its particular solutions.

\section{Unfolded dynamics approach and classical scalar field\label{SEC_UDA}}

\subsection{General construction}

In the unfolded dynamics approach \citep{unf1,vas1,vas2,unf2,ActionsCharges},
a classical field theory is formulated through imposing unfolded equations
\begin{equation}
\mathrm{d}W^{A}(x)+G^{A}(W)=0\label{unf_eq}
\end{equation}
on unfolded fields $W^{A}(x)$, where $A$ stands for all indices
of the field. The theory is formulated on some space-time manifold
$M^{d}$ with local coordinates $x$ and de Rham differential $\mathrm{d}$.
Unfolded fields $W$ are exterior forms on $M^{d}$, and $G^{A}(W)$
is built from exterior products of $W$ (we omit the wedge symbol
throughout the paper). Every unfolded field $W^{A}$ is provided with
one and only one own unfolded equation \eqref{unf_eq}.

The nilpotency of the de Rham differential $\mathrm{d}^{2}\equiv0$
entails a consistency condition for $G$ 
\begin{equation}
G^{B}\dfrac{\delta G^{A}}{\delta W^{B}}\equiv0,\label{unf_consist}
\end{equation}
which plays a central role in the process of \textquotedbl unfolding\textquotedbl{}
the field theory. Different solutions to \eqref{unf_consist} can,
in general, provide different unfolded formulations for the same field
theory.

If \eqref{unf_consist} holds, unfolded equations \eqref{unf_eq}
are manifestly invariant under infinitesimal gauge transformations
\begin{equation}
\delta W^{A}=\mathrm{d}\varepsilon^{A}(x)-\varepsilon^{B}\dfrac{\delta G^{A}}{\delta W^{B}}.\label{unf_gauge_transf}
\end{equation}
Here a gauge parameter $\varepsilon^{A}(x)$, representing a rank-$(n-1)$
form, is generated by a rank-$n$ unfolded field $W^{A}$. 0-form
unfolded fields do not generate gauge symmetries and are transformed
only by gauge symmetries of higher-rank fields through the second
term in \eqref{unf_gauge_transf}. At the linear level, 0-forms are
transformed only due to vacuum symmetries and therefore correspond
to gauge-invariant degrees of freedom. In a nutshell, an unfolded
field includes some physical field (we call it a primary field) and
all its differential descendants, parameterized in a coordinate-independent
way. In a nonlinear theory, the basis of differential descendants
usually becomes nonlinear as well.

The two most important features of the unfolded dynamics approach
are the manifest gauge invariance, which allows one to efficiently
control all gauge symmetries of a theory, and the manifest coordinate
independence, ensured by the exterior form formalism.

\subsection{Unfolded Minkowski vacuum}

According to the ideology of the unfolded dynamics approach, the geometry
of the space-time manifold $M^{d}$ must be encoded in some unfolded
equations \eqref{unf_eq}. This is achieved by using the Cartan formalism.

One introduces a 1-form connection $\Omega=\mathrm{d}x^{\underline{a}}\Omega_{\underline{a}}^{A}(x)T_{A}$
that takes values in the Lie algebra of symmetries of $M^{d}$ with
generators $T_{A}$. Then a maximally symmetric vacuum arises via
imposing Maurer--Cartan equation on $\Omega$ 
\begin{equation}
\mathrm{d}\Omega+\frac{1}{2}[\Omega,\Omega]=0\label{flat_conn}
\end{equation}
(square brackets stand for the Lie-algebra commutator). Fixing some
particular solution $\Omega_{0}$ to this equation breaks the associated
gauge symmetry \eqref{unf_gauge_transf} 
\begin{equation}
\delta\Omega=\mathrm{d}\varepsilon(x)+[\Omega,\varepsilon]
\end{equation}
down to a residual global symmetry $\varepsilon_{glob}$, that leaves
the solution $\Omega_{0}$ invariant and thus must satisfy 
\begin{equation}
\mathrm{d}\varepsilon_{glob}+[\Omega_{0},\varepsilon_{glob}]=0.\label{glob_symm}
\end{equation}
In the paper we deal with $4d$ Minkowski space, so we consider a
connection that takes values in Poincaré algebra $iso(1,3)$ 
\begin{equation}
\Omega=e^{\alpha\dot{\beta}}P_{\alpha\dot{\beta}}+\omega^{\alpha\beta}M_{\alpha\beta}+\bar{\omega}^{\dot{\alpha}\dot{\beta}}\bar{M}_{\dot{\alpha}\dot{\beta}},\label{AdS_connection}
\end{equation}
where $P_{\alpha\dot{\alpha}}$, $M_{\alpha\beta}$ and $\bar{M}_{\dot{\alpha}\dot{\beta}}$
represent generators of space-time translations and (selfdual and
anti-selfdual part of) rotations. Fields $e^{\alpha\dot{\beta}}$
and $\omega^{\alpha\beta}$ ($\bar{\omega}^{\dot{\alpha}\dot{\beta}}$)
are 1-forms of vierbein and Lorentz connection. Two-valued indices
$\alpha$ and $\dot{\beta}$ correspond to two spinor representations
of the Lorentz algebra $so(3,1)\approx sl(2,\mathbb{C})$. The indices
are raised and lowered by the Lorentz-invariant spinor metric 
\begin{equation}
\epsilon_{\alpha\beta}=\epsilon_{\dot{\alpha}\dot{\beta}}=\left(\begin{array}{cc}
0 & 1\\
-1 & 0
\end{array}\right),\quad\epsilon^{\alpha\beta}=\epsilon^{\dot{\alpha}\dot{\beta}}=\left(\begin{array}{cc}
0 & 1\\
-1 & 0
\end{array}\right)
\end{equation}
according to 
\begin{equation}
v_{\alpha}=\epsilon_{\beta\alpha}v^{\beta},\quad v^{\alpha}=\epsilon^{\alpha\beta}v_{\beta},\quad\bar{v}_{\dot{\alpha}}=\epsilon_{\dot{\beta}\dot{\alpha}}\bar{v}^{\dot{\beta}},\quad\bar{v}^{\dot{\alpha}}=\epsilon^{\dot{\alpha}\dot{\beta}}\bar{v}_{\dot{\beta}}.
\end{equation}

The expansion of \eqref{flat_conn} in terms of generators gives 
\begin{align}
 & \mathrm{d}e^{\alpha\dot{\beta}}+\omega^{\alpha}\text{}_{\gamma}e^{\gamma\dot{\beta}}+\bar{\omega}^{\dot{\beta}}\text{}_{\dot{\gamma}}e^{\alpha\dot{\gamma}}=0,\label{mink1}\\
 & \mathrm{d}\omega^{\alpha\beta}+\omega^{\alpha}\text{}_{\gamma}\omega^{\gamma\beta}=0,\\
 & \mathrm{d}\bar{\omega}^{\dot{\alpha}\dot{\beta}}+\bar{\omega}^{\dot{\alpha}}\text{}_{\dot{\gamma}}\bar{\omega}^{\dot{\gamma}\dot{\beta}}=0.\label{mink2}
\end{align}
The simplest solution to \eqref{mink1}-\eqref{mink2} (with a non-degenerate
vierbein) is provided by Cartesian coordinates 
\begin{equation}
e_{\underline{m}}{}^{\alpha\dot{\beta}}=(\bar{\sigma}_{\underline{m}})^{\dot{\beta}\alpha},\quad\omega_{\underline{m}}{}^{\alpha\beta}=0,\quad\bar{\omega}_{\underline{m}}{}^{\dot{\alpha}\dot{\beta}}=0.\label{cartes_coord}
\end{equation}
In these coordinates, equation \eqref{glob_symm} is solved by 
\begin{equation}
\varepsilon_{glob}^{\alpha\dot{\beta}}=\xi^{\alpha\dot{\beta}}+\xi^{\alpha}\text{}_{\gamma}(\bar{\sigma}_{\underline{m}})^{\dot{\beta}\gamma}x^{\underline{m}}+\bar{\xi}^{\dot{\beta}}\text{}_{\dot{\gamma}}(\bar{\sigma}_{\underline{m}})^{\dot{\gamma}\alpha}x^{\underline{m}},\quad\varepsilon_{glob}^{\alpha\beta}=\xi^{\alpha\beta},\quad\bar{\varepsilon}_{glob}{}^{\dot{\alpha}\dot{\beta}}=\bar{\xi}^{\dot{\alpha}\dot{\beta}}\label{glob_cart_sol}
\end{equation}
with $x$-independent $\xi^{\alpha\dot{\beta}}$, $\xi^{\alpha\beta}$
and $\bar{\xi}^{\dot{\alpha}\dot{\beta}}$ being parameters of global
Poincaré transformations.

When we consider an unfolded scalar field in the next Subsection,
\eqref{unf_gauge_transf} with $\varepsilon_{glob}$ will define a
representation of Poincaré algebra on the unfolded module.

\subsection{Unfolded self-interacting scalar field\label{SUB_CLASS_SCALAR}}

Unfolded formulation of the scalar field $\phi(x)$ requires the introduction
of an infinite sequence of 0-forms, which, as we will see, encode
all its linearly independent differential descendants.

We start with defining an unfolded scalar field as the following set
of 0-forms 
\begin{equation}
\Phi(Y,\tau|x)=\sum_{n=0}^{\infty}\sum_{k=0}^{\infty}\Phi_{n}^{(k)}(Y,\tau|x)=\sum_{n=0}^{\infty}\sum_{k=0}^{\infty}\frac{1}{(n!)^{2}}\Phi_{\alpha(n),\dot{\alpha}(n)}^{(k)}(x)y^{\alpha_{1}}...y^{\alpha_{n}}\bar{y}^{\dot{\alpha}_{1}}...\bar{y}^{\dot{\alpha}_{n}}\frac{\tau^{k}}{k!}.\label{F_scalar}
\end{equation}
Here we make use of the condensed notations for symmetric spinor-tensors,
so that 
\begin{equation}
f_{\alpha(n)}:=f_{\alpha_{1}...\alpha_{n}}.
\end{equation}
Contracting all spinor indices of $\Phi_{\alpha(n),\dot{\alpha}(n)}$
with $\sigma$-matrices, one can see that it corresponds to a symmetric
traceless rank-$n$ Lorentz tensor 
\begin{equation}
\Phi_{a_{1}a_{2}...a_{n}}=(\bar{\sigma}_{a_{1}})^{\dot{\alpha}_{1}\alpha_{1}}...(\bar{\sigma}_{a_{n}})^{\dot{\alpha}_{n}\alpha_{n}}\Phi_{\alpha(n),\dot{\alpha}(n)},\quad\eta^{a_{1}a_{2}}\Phi_{a_{1}a_{2}...a_{n}}=0.
\end{equation}
Thus, \eqref{F_scalar} is equivalent to a set of symmetric traceless
Lorentz tensors of all ranks, dependent on space-time coordinate $x^{\underline{m}}$
and additional variables $y^{\alpha}$, $\bar{y}^{\dot{\alpha}}$,
$\tau$.

A pair of auxiliary commuting $sl(2,\mathbb{C})$-spinors $Y=(y^{\alpha},\bar{y}^{\dot{\alpha}})$
in \eqref{F_scalar} is introduced for the convenience of operating
with symmetric spinor-tensors. Due to their commutativity, $Y$ are
null with respect to the antisymmetric spinor metric 
\begin{equation}
y^{\alpha}y^{\beta}\epsilon_{\alpha\beta}=0,\quad\bar{y}^{\dot{\alpha}}\bar{y}^{\dot{\beta}}\epsilon_{\dot{\alpha}\dot{\beta}}=0.\label{null_Y}
\end{equation}
We also define corresponding derivatives as 
\begin{equation}
\partial_{\alpha}y^{\beta}=\delta_{\alpha}\text{}^{\beta},\quad\bar{\partial}_{\dot{\alpha}}\bar{y}^{\dot{\beta}}=\delta_{\dot{\alpha}}\text{}^{\dot{\beta}}
\end{equation}
and an Euler operator $N$ 
\begin{equation}
N=\frac{1}{2}y^{\alpha}\partial_{\alpha}+\frac{1}{2}\bar{y}^{\dot{\alpha}}\bar{\partial}_{\dot{\alpha}}.\label{Euler}
\end{equation}
As becomes clear below, higher powers in $y\bar{y}$ and $\tau$ in
\eqref{F_scalar} correspond to differential descendants of a scalar
field 
\begin{equation}
\phi(x)=\Phi(Y=0,\tau=0|x),\label{primary_phi}
\end{equation}
which for this reason we call the primary field.

A consistent unfolded system, that describes a self-interacting primary
scalar $\phi$, is 
\begin{equation}
\mathrm{D}\Phi-\frac{1}{N+1}e^{\alpha\dot{\beta}}\partial_{\alpha}\bar{\partial}_{\dot{\beta}}\Phi+\frac{1}{N+1}e^{\alpha\dot{\beta}}y_{\alpha}\bar{y}_{\dot{\beta}}\left(m^{2}\Phi+g\mathrm{U}'(\Phi)-\frac{\partial}{\partial\tau}\Phi\right)=0,\label{F_eq}
\end{equation}
where $\mathrm{U}'$ corresponds to the first variation of the scalar
potential, $g$ is a coupling constant and $\mathrm{D}$ is the Lorentz-covariant
derivative 
\begin{equation}
\mathrm{D}f(Y,\tau|x):=\left(\mathrm{d}+\omega^{\alpha\beta}y_{\alpha}\partial_{\beta}+\bar{\omega}^{\dot{\alpha}\dot{\beta}}\bar{y}_{\dot{\alpha}}\bar{\partial}_{\dot{\beta}}\right)f(Y,\tau|x),
\end{equation}
which in Cartesian coordinates comes down to the de Rham differential.
A family of all unfolded realizations of this model, generalizing
\eqref{F_eq}, is discussed in Appendix A.

Let us analyze the content of \eqref{F_eq}. To this order we expand
the Lorentz-covariant derivative in the vierbein as
\begin{equation}
\mathrm{D}=e^{\alpha\dot{\alpha}}\nabla_{\alpha\dot{\alpha}}
\end{equation}
and act on \eqref{F_eq} with
\begin{equation}
y^{\alpha}\bar{y}^{\dot{\beta}}\frac{\delta}{\delta e^{\alpha\dot{\beta}}},
\end{equation}
which yields a relation, that completely determines $Y$-dependence
of $\Phi$,
\begin{equation}
\left(y^{\alpha}\bar{y}^{\dot{\alpha}}\nabla_{\alpha\dot{\alpha}}-N\right)\Phi=0.\label{Y_eq}
\end{equation}
We see that the whole $Y$-dependence in fact arises as a simple shift
of $x$ in $\Phi$ by $y^{\alpha}\bar{y}^{\dot{\alpha}}$, 
\begin{equation}
\Phi(Y,\tau|x)=\exp\left(y^{\alpha}\bar{y}^{\dot{\alpha}}\nabla_{\alpha\dot{\alpha}}\right)\Phi(0,\tau|x),\label{Y_shift}
\end{equation}
or, treated another way, $y\bar{y}$ parameterize all traceless (because
of \eqref{null_Y}) derivatives of $\Phi(0,\tau|x)$.

To determine $\tau$-dependence of $\Phi$, we act on \eqref{F_eq}
with 
\begin{equation}
(\nabla^{\alpha\dot{\alpha}}+\frac{1}{N+1}\partial^{\alpha}\bar{\partial}^{\dot{\alpha}})\frac{\delta}{\delta e^{\alpha\dot{\alpha}}},
\end{equation}
which, accounting for \eqref{Y_eq}, leads to 
\begin{equation}
\square\Phi+m^{2}\Phi+g\mathrm{U}'(\Phi)=\dot{\Phi},\label{tau_eq}
\end{equation}
where the dot stands for the $\tau$-derivative and d'Alembertian
is defined as 
\begin{equation}
\square:=\frac{1}{2}\nabla_{\alpha\dot{\alpha}}\nabla^{\alpha\dot{\alpha}}.
\end{equation}
Making use of \eqref{Y_shift}, we deduce hereof 
\begin{equation}
(\square+m^{2})\Phi(0,\tau|x)+g\mathrm{U}'(\Phi(0,\tau|x))=\dot{\Phi}(0,\tau|x),\label{tau_eq_no_Y}
\end{equation}
This equation fixes $\tau$-dependence of $\Phi$. Of course, for
general $\mathrm{U}'$ this equation cannot be resolved manifestly,
as opposite to the $Y$-equation \eqref{Y_eq}, so the dependence
on $\tau$ can be very complicated.

But in the case of the free theory $\mathrm{U}'=0$, a manifest solution
to 
\begin{equation}
(\square+m^{2})\Phi=\dot{\Phi}\label{Phi_free_tau_eq}
\end{equation}
can be easily written as 
\begin{equation}
\Phi^{free}(0,\tau|x)=\exp\left(\tau(\square+m^{2})\right)\phi(x)\label{tau_shift}
\end{equation}
where $\phi$ is the primary scalar field \eqref{primary_phi}. Combining
\eqref{tau_shift} with \eqref{Y_shift}, one finds the full solution
to the unfolded system \eqref{F_eq} with $\mathrm{U}'=0$ to be 
\begin{equation}
\Phi^{free}(Y,\tau|x)=e^{\tau(\square+m^{2})+y^{\alpha}\bar{y}^{\dot{\alpha}}\nabla_{\alpha\dot{\alpha}}}\phi(x).\label{Phi_free}
\end{equation}
Now one can give a clear interpretation of $Y$- and $\tau$-dependent
components of the free unfolded field $\Phi$: they provide a basis
in the space of differential descendants of the primary scalar $\phi$,
with $y\bar{y}$ parameterizing traceless derivatives and $\tau$
parameterizing powers of the kinetic operator $(\square+m^{2})$.
These two sequences exhaust all possible types of descendants in the
case of the scalar field.

In the case of general $\mathrm{U}'\neq0$, one can write down a formal
implicit solution as 
\begin{equation}
\Phi(0,\tau|x)=e^{\tau(\square+m^{2})}\phi(x)+g\intop_{0}^{\tau}d\tau'e^{(\tau-\tau')(\square+m^{2})}\mathrm{U}'(\Phi(0,\tau'|x)).\label{Phi_inter}
\end{equation}
Here the first term coincides with the free solution \eqref{tau_shift},
so one can solve \eqref{Phi_inter} perturbatively in $g$. Comparing
\eqref{Phi_inter} with \eqref{Phi_free}, we see that $\tau$-dependence
of the self-interacting $\Phi$ plays the same role as in the free
case: it encodes d'Alembertians of $\phi$, but now sophisticatedly
entangled with nonlinear corrections coming from the potential. Strictly
speaking, all systems \eqref{F_eq} with different $\mathrm{U}'$
are in certain sense equivalent, as they simply provide different
parameterizations for the space of differential descendants of $\phi$.
However, this equivalence is established, in general, by strongly
non-local and non-linear field redefinitions.

The unfolded system \eqref{F_eq} is said to be off-shell, because
the primary field $\phi(x)$ is not subjected to any differential
constraints like e.g. equations of motion. To put the system on-shell,
i.e. to subject $\phi$ to some differential constraints, one has
to consistently remove some part of descendants inside of $\Phi$.
An advantage of the system \eqref{F_eq} is that the on-shell reduction,
which leads to the standard e.o.m. for the self-interacting scalar
with potential $\mathrm{U}(\phi)$, is realized by a simple constraint
\begin{equation}
\dot{\Phi}=0,\label{d_tau_Phi_zero}
\end{equation}
which eliminates all $\tau$-descendants from $\Phi$.

To simplify notations, from now on we omit spinor indices, contracted
between a vierbein 1-form $e^{\alpha\dot{\beta}}$ and auxiliary spinors,
and write 
\begin{equation}
ey\bar{y}:=e^{\alpha\dot{\beta}}y_{\alpha}\bar{y}_{\dot{\beta}},\quad e\partial\bar{\partial}:=e^{\alpha\dot{\beta}}\partial_{\alpha}\bar{\partial}_{\dot{\beta}}.
\end{equation}

Then an unfolded equation for $\Phi=\Phi(Y|x)$ becomes 
\begin{equation}
\mathrm{D}\Phi-\frac{1}{N+1}e\partial\bar{\partial}\Phi+\frac{1}{N+1}ey\bar{y}\left(m^{2}\Phi+g\mathrm{U}'(\Phi)\right)=0\label{F_eq_on_shell}
\end{equation}
and imposes, as follows from \eqref{tau_eq}, a differential constraint
\begin{equation}
(\square+m^{2})\Phi+g\mathrm{U}'(\Phi)=0,
\end{equation}
which includes, at $Y=0$, e.o.m. for the primary scalar 
\begin{equation}
(\square+m^{2})\phi+g\mathrm{U}'(\phi)=0.\label{KG_eq}
\end{equation}
So in this case the unfolded field $\Phi$ describes the primary scalar
$\phi$, subjected to nonlinear Klein--Gordon equation \eqref{KG_eq},
and all its independent non-zero descendants encoded in $Y$-expansion
\eqref{Y_shift} (note that the on-shell reduction \eqref{d_tau_Phi_zero}
does not affect $Y$-sector of the problem), 
\begin{equation}
\Phi^{on\textrm{-}shell}(Y|x)=\exp\left(y^{\alpha}\bar{y}^{\dot{\alpha}}\nabla_{\alpha\dot{\alpha}}\right)\phi(x).
\end{equation}

Now let us return to the off-shell system \eqref{F_eq}. Keeping in
mind the form of the on-shell constraint \eqref{d_tau_Phi_zero},
we see that $\dot{\Phi}(Y=0,\tau=0|x)$ can be treated as an external
source $j(x)$ for the primary scalar $\phi(x)$. This follows from
\eqref{tau_eq}, projected to ($Y=0,$ $\tau=0$)-component, 
\begin{equation}
(\square+m^{2})\phi+g\mathrm{U}'(\phi)=j(x),\quad j(x):=\dot{\Phi}(Y=0,\tau=0|x).\label{phi_j_eq}
\end{equation}
Thus, an off-shell unfolded model can be as well treated as the on-shell
one, coupled to an external source \citep{misuna}. This observation
plays a decisive role when one turns to the problem of quantization
of an unfolded theory.

\subsection{Relation to Vasiliev Higher-Spin Gravity}

Let us take a cursory glance at how the pieces of the unfolding formalism
we considered so far are built into Vasiliev's unfolded formulation
of $4d$ on-shell HS gravity \citep{vas1,vas2}.

First, the space of 0-forms of the higher-spin theory contains spinor-tensors
of all possible ranks in dotted and undotted indices $C_{\alpha(m),\dot{\beta}(n)}$,
not only a scalar unfolded module $\Phi_{\alpha(n),\dot{\alpha}(n)}$.
These new 0-forms correspond to gauge-invariant strength tensors of
all fields of the theory (Maxwell tensor and its descendants for $s=1$,
Weyl tensor and its descendants for $s=2$ and so on for higher-spin
fields). Analogously, 1-forms of the theory now include all possible
$\omega_{\alpha(m),\dot{\beta}(n)}$ besides the gravitational sector
$m+n=2$, which we have used to describe Minkowski vacuum. They encode
potentials of gauge fields and their gauge-non-invariant descendants
(first $(s-1)$ derivatives of the potential for a spin-$s$ field).
At the linear order, 1-forms get connected to corresponding 0-forms,
that makes them dynamical (in particular, the gravitational gauge
multiplet $e^{\alpha\dot{\beta}}$, $\omega^{\alpha\beta}$, $\bar{\omega}^{\dot{\alpha}\dot{\beta}}$
gets connected to 0-forms of Weyl tensor $C_{\alpha(4)}$, $\bar{C}_{\dot{\alpha}(4)}$,
that allows a metric to fluctuate). At the higher orders both 0- and
1-form equations receive nonlinear corrections that describe HS interactions.

All spinor indices are still contracted with $Y$-spinors, which now
play the very important role, being generating elements of infinite-dimensional
associative algebra of HS gauge symmetries.

Finally, Minkowski vacuum \eqref{mink1}-\eqref{mink2} is not a solution
of Vasiliev theory. HS gauge symmetry requires a non-zero value of
the cosmological constant, so usually one considers an expansion over
$AdS_{4}$ background.

For a detailed review of Vasiliev theory see e.g. \citep{Vas_StarProduct,elem_vas}.

\section{Quantization of the unfolded scalar field\label{SEC_SD}}

In this Section, we develop a method of quantization of the classical
unfolded system presented in the previous Section. In short, making
use of the analogy with functional Schwinger--Dyson equations, we
promote off-shell unfolded equations 
\[
\mathrm{d}W^{A}+G^{A}(W)=0
\]
to operator equations for a partition function $Z$ 
\[
\left(\mathrm{d}\hat{W}^{A}+\hat{G}^{A}(\hat{W})\right)Z=0
\]
and determine an unfolded operator algebra 
\[
[\hat{W}^{A},\hat{W}^{B}]=F^{A,B}(\hat{W}).
\]

\subsection{Functional Schwinger--Dyson equation\label{SUB_FUNCT_SD}}

In a standard QFT with the classical action $S[\phi]$, the partition
function 
\begin{equation}
Z[j]:=\int\mathcal{D}\phi\exp\{\frac{i}{\hbar}S[\phi]-\frac{i}{\hbar}\int d^{4}x\phi(x)j(x)\}
\end{equation}
satisfies the functional Schwinger--Dyson equation 
\begin{equation}
\frac{\delta S}{\delta\phi}[i\hbar\frac{\delta}{\delta j}]Z=jZ,\label{SD_eq}
\end{equation}
which can be deduced from the fact that a functional integral of a
total derivative vanishes, so that 
\begin{equation}
\int\mathcal{D}\phi\frac{\delta}{\delta\phi}e^{\frac{i}{\hbar}(S-\int d^{4}x\phi j)}=0.
\end{equation}

Schwinger--Dyson equation \eqref{SD_eq} can be obtained as follows.
One starts with the classical e.o.m. of the theory coupled to an external
source $j$, 
\begin{equation}
\frac{\delta S}{\delta\phi}[\phi]=j,\label{class_eq}
\end{equation}
and \textquotedbl quantize\textquotedbl{} it by promoting a field-source
pair to the operators acting on a \textquotedbl wave function\textquotedbl{}
$Z$ and obeying canonical commutation relations 
\begin{equation}
[\hat{\phi}(x_{1}),\hat{j}(x_{2})]=i\hbar\delta^{4}(x_{1}-x_{2}),\quad[\hat{\phi}(x_{1}),\hat{\phi}(x_{2})]=[\hat{j}(x_{1}),\hat{j}(x_{2})]=0.\label{primary_comm}
\end{equation}
Then, in $j$-representation, one arrives at \eqref{SD_eq}. It should
be stressed that the first-quantized system \eqref{primary_comm}
has no relation to usual canonical commutation relations of a second-quantized
field theory (where, of course, operators of a quantum field do not
commute), and represents just a formal trick, introduced in \citep{Anchor2},
that allows one to arrive at \eqref{SD_eq} starting from \eqref{class_eq}.

We are going to perform a similar procedure for the unfolded system
\eqref{F_eq}. First, we consider the simpler case of the free theory,
$\mathrm{U}'(\Phi)=0$.

\subsection{Free quantum scalar\label{SUB_FREE_QSCALAR}}

We start with the classical off-shell unfolded equation 
\begin{equation}
(N+1)\mathrm{D}\Phi-e\partial\bar{\partial}\Phi+ey\bar{y}(m^{2}\Phi-\dot{\Phi})=0.\label{free_class_eq}
\end{equation}
Accounting for \eqref{phi_j_eq}, we define an unfolded external source
as 
\begin{equation}
\mathrm{J}(Y,\tau|x):=\dot{\Phi}(Y,\tau|x)\label{Legendre}
\end{equation}
(remarkably, this formally looks like a transition from a \textquotedbl velocity\textquotedbl{}
$\dot{\Phi}$ to a conjugate \textquotedbl momentum\textquotedbl{}
$\mathrm{J}$).

Now we have a pair of unfolded equations

\begin{equation}
(N+1)\mathrm{D}\Phi-e\partial\bar{\partial}\Phi+ey\bar{y}(m^{2}\Phi-\mathrm{J})=0,\label{Phi_J_eq}
\end{equation}
\begin{equation}
(N+1)\mathrm{D}\mathrm{J}-e\partial\bar{\partial}\mathrm{J}+ey\bar{y}(m^{2}\mathrm{J}-\dot{\mathrm{J}})=0.\label{J_eq}
\end{equation}
Analogously to \eqref{Phi_free}, a solution to \eqref{J_eq} is 
\begin{equation}
\mathrm{J}(Y,\tau|x)=\exp\left(\tau(\square+m^{2})+y^{\alpha}\bar{y}^{\dot{\alpha}}\nabla_{\alpha\dot{\alpha}}\right)j(x),\label{J_sol}
\end{equation}
where $j(x)$ is a primary source. Instead of \eqref{Phi_free_tau_eq},
a wave equation for $\Phi$ is now 
\begin{equation}
(\square+m^{2})\Phi=\mathrm{J},\label{Phi_J_wave}
\end{equation}
and in the primary ($Y=0,$ $\tau=0$)-sector

\begin{equation}
(\square+m^{2})\phi=j.\label{phi_j_wave}
\end{equation}
Treating \eqref{phi_j_wave} as a classical e.o.m. with an external
source $j$, the theory can be quantized as described in Subsection
\ref{SUB_FUNCT_SD}. However, our goal is to get a closed formulation
of the unfolded quantum theory in terms of the unfolded quantum fields
$\hat{\Phi}$ and $\hat{\mathrm{J}}$, without manifest appealing
to primary fields, which are just their particular components.

We want to promote the system \eqref{Phi_J_eq}-\eqref{J_eq} to the
quantum operator equations on the partition function $Z$ 
\begin{equation}
\left((N+1)\mathrm{D}\hat{\Phi}-e\partial\bar{\partial}\hat{\Phi}+ey\bar{y}(m^{2}\hat{\Phi}-\hat{\mathrm{J}})\right)Z=0,\label{Phi_J_qeq}
\end{equation}
\begin{equation}
\left((N+1)\mathrm{D}-e\partial\bar{\partial}+ey\bar{y}(m^{2}-\frac{\partial}{\partial\tau})\right)\hat{\mathrm{J}}Z=0.\label{J_qeq}
\end{equation}
This requires the definition of the commutator $[\hat{\Phi}_{i},\hat{\mathrm{J}}_{k}]$,
satisfying an \textquotedbl initial condition\textquotedbl{} \eqref{primary_comm}
and consistent with \eqref{Phi_J_qeq}-\eqref{J_qeq}. From now on,
a subscript of an unfolded field denotes the full set of its arguments,
i.e. 
\begin{equation}
\Phi_{i}:=\Phi(Y_{i},\tau_{i}|x_{i}).
\end{equation}

Resolving $Y$-dependence in \eqref{Phi_J_qeq}-\eqref{J_qeq} (which
still comes down to a shift of space-time coordinates $x$ by $y\bar{y}$,
as in the classical theory), one extracts wave equations 
\begin{equation}
(\square\hat{\Phi}+m^{2}\hat{\Phi}-\hat{\mathrm{J}})Z=0,\label{free_Z_wave_eq}
\end{equation}
\begin{equation}
(\square+m^{2}-\frac{\partial}{\partial\tau})\hat{\mathrm{J}}Z=0.\label{J_free_Z}
\end{equation}
Assuming naturally 
\begin{equation}
[\hat{\Phi}_{i},\hat{\Phi}_{k}]=[\hat{\mathrm{J}}_{i},\hat{\mathrm{J}}_{k}]=0,\label{F_F_comm_J_J_comm}
\end{equation}
a self-consistency of these equations requires 
\begin{equation}
(\square_{i}+m^{2})[\hat{\Phi}_{i},\hat{\mathrm{J}}_{k}]=(\square_{k}+m^{2})[\hat{\Phi}_{k},\hat{\mathrm{J}}_{i}],\label{qcons_1_free}
\end{equation}
\begin{equation}
(\square_{i}+m^{2})(\square_{k}+m^{2}-\frac{\partial}{\partial\tau_{k}})[\hat{\Phi}_{i},\hat{\mathrm{J}}_{k}]=0.\label{qcons_2_free}
\end{equation}
Any solution to this system, respecting the initial condition \eqref{primary_comm},
defines some consistent quantization of the unfolded system \eqref{free_class_eq}.

We pick up a particular solution, which in Cartesian coordinates \eqref{cartes_coord}
is 
\begin{equation}
[\hat{\Phi}_{i},\hat{\mathrm{J}}_{k}]=i\hbar K_{\tau_{i}+\tau_{k}}^{m}(x_{i}+y_{i}\bar{y}_{i};x_{k}+y_{k}\bar{y}_{k}),\label{F_J_comm_free}
\end{equation}
where 
\begin{equation}
K_{\tau}^{m}(x_{1};x_{2}):=e^{m^{2}\tau}K_{\tau}(x_{1};x_{2})
\end{equation}
and the heat kernel $K_{\tau}(x_{1};x_{2})$ is defined in the usual
way 
\begin{equation}
K_{\tau}(x_{1};x_{2}):=\frac{1}{(4\pi\tau)^{2}}\exp\left\{ -\frac{(x_{1}-x_{2})^{2}}{2\tau}\right\} 
\end{equation}
and possesses well-known properties 
\begin{equation}
(\square_{i=\{1,2\}}+m^{2}-\frac{\partial}{\partial\tau})e^{m^{2}\tau}K_{\tau}(x_{1};x_{2})=0,\label{diff_eq_K}
\end{equation}
\begin{equation}
\underset{\tau\rightarrow0}{\lim}K_{\tau}(x_{1};x_{2})=\delta(x_{1}-x_{2}),\label{tau_0_K}
\end{equation}
\begin{equation}
\int d^{4}x_{2}K_{\tau}(x_{1};x_{2})K_{\tau'}(x_{2};x_{3})=K_{\tau+\tau'}(x_{1};x_{3}).
\end{equation}
The commutator \eqref{F_J_comm_free} satisfies \eqref{qcons_1_free}-\eqref{qcons_2_free}
due to \eqref{diff_eq_K} and symmetricity with respect to $\tau_{i}\leftrightarrow\tau_{k}$,
while \eqref{tau_0_K} guarantees that \eqref{primary_comm} holds.

The expression for the commutator of unfolded quantum fields \eqref{F_J_comm_free}
is one of the central ones in the paper. Let us pay attention to two
of its most remarkable features:

(1) It naturally contains the heat kernel in the Schwinger proper-time
parameterization. However, while the proper time $\tau$ in Schwinger's
method appears as a formal integration variable which allows for the
convenient representations of Green's functions, one-loop determinants
etc., in the unfolded dynamics approach it arises already at the classical
level and possesses a clear interpretation -- $\tau$ parameterizes
off-shell descendants of the primary field and generates the transform
\eqref{Legendre}, which defines an unfolded source conjugate to the
unfolded field;

(2) Expression \eqref{F_J_comm_free} does not immediately produce
a singularity in coinciding space-time points. A singularity develops
only when $\tau_{i}=\tau_{k}=0$, i.e. in the sector of primary fields
and their on-shell descendants. Effectively, the $\tau$-dependent
heat kernel replaces for unfolded fields the space-time delta-function
of a standard QFT. Thus, the proper time $\tau$ serves as a natural
regularizer, which potentially might manage and soften quantum divergences;

Finally, \eqref{F_J_comm_free} is to some extent similar to the propagator
of a non-relativistic quantum particle, as usual for the heat kernels.
An important difference, however, is that \eqref{F_J_comm_free} depends
on the sum of proper times, not on the difference.

Now let us use \eqref{F_J_comm_free} to complete the quantization
of the free unfolded scalar. We go to $\mathrm{J}$-representation,
where unfolded operators are realized, according to \eqref{F_J_comm_free},
as 
\begin{equation}
\hat{\mathrm{J}}_{i}=\mathrm{J}(Y_{i},\tau_{i}|x_{i})
\end{equation}
\begin{equation}
\hat{\Phi}_{i}=i\hbar\int d^{4}x'K_{\tau_{i}+\tau'}^{m}(x_{i}+y\bar{y}_{i};x'+y\bar{y}')\frac{\delta}{\delta\mathrm{J}(Y',\tau'|x')}.\label{Phi_op_in J_rep}
\end{equation}
In $\mathrm{J}$-representation, \eqref{J_qeq} coincides with the
classical equation \eqref{J_eq}, whose solution is \eqref{J_sol}.
The variation $\frac{\delta}{\delta\mathrm{J}}$ in \eqref{Phi_op_in J_rep}
is defined as 
\begin{equation}
\frac{\delta}{\delta\mathrm{J}_{k}}\mathrm{J}_{i}=K_{\tau_{i}-\tau_{k}}^{m}(x_{i}+y\bar{y}_{i};x_{k}+y\bar{y}_{k}),
\end{equation}
which corresponds to the free variation of the primary source 
\begin{equation}
\frac{\delta}{\delta j(x_{k})}j(x_{i})=\delta(x_{i}-x_{k}),
\end{equation}
followed by the unfolding map \eqref{J_sol}.

When considering a free theory, it is more convenient to deal with
a generator for connected correlation functions $W$ instead of $Z$,
\begin{equation}
W=\log Z.\label{W_def}
\end{equation}
Then \eqref{free_Z_wave_eq} turns to 
\begin{equation}
(\square+m^{2})i\hbar\int d^{4}x'K_{\tau+\tau'}^{m}(x+y\bar{y};x'+y\bar{y}')\frac{\delta W}{\delta\mathrm{J}(Y',\tau'|x')}=\mathrm{J}(Y,\tau|x),
\end{equation}
which can be solved as 
\begin{equation}
W[\mathrm{J}]=-\frac{i}{2\hbar}\int d^{4}x\mathrm{J}(\tau=0,Y|x)\left(\square+m^{2}\right)^{-1}\mathrm{J}(\tau=0,Y|x),\label{W_J_free}
\end{equation}
if one takes into account \eqref{J_sol}. Substituting \eqref{J_sol}
into \eqref{W_J_free}, one can express $W$ solely in terms of primary
sources as 
\begin{equation}
W[j]=-\frac{i}{2\hbar}\int d^{4}xj(x)\left(\square+m^{2}\right)^{-1}j(x),\label{W_usual}
\end{equation}
which is standard $W$ of a free scalar.

Using \eqref{W_J_free}, one recovers a propagator of the unfolded
free scalar field 
\begin{equation}
\bigl\langle\Phi_{i}\Phi_{k}\bigr\rangle^{0}=\hat{\Phi}_{i}\hat{\Phi}_{k}W|_{\mathrm{J}=0}=\frac{i\hbar}{\square_{i}+m^{2}}K_{\tau_{i}+\tau_{k}}^{m}(x_{i}+y\bar{y}_{i};x_{k}+y\bar{y}_{k}).\label{unf_free_prop}
\end{equation}
Again, we see that the space-time delta-function is replaced by the
heat kernel.

In the primary sector, one has, sending $\tau$ and $Y$ to zero and
using \eqref{tau_0_K}, 
\begin{equation}
\bigl\langle\phi(x_{i})\phi(x_{k})\bigr\rangle^{0}=\frac{i\hbar}{\square_{i}+m^{2}}\delta(x_{i}-x_{k}),
\end{equation}
i.e. the standard propagator of a free scalar.

Finally, let us formulate the unfolded operator algebra \eqref{F_F_comm_J_J_comm},
\eqref{F_J_comm_free} in a coordinate-independent way, without appealing
to the Cartesian frame. To this end, we treat the commutator 
\begin{equation}
\hat{\mathrm{C}}_{ik}^{0}=[\hat{\Phi}_{i},\hat{\mathrm{J}}_{k}]
\end{equation}
as a new two-point unfolded field with its own unfolded equations
and certain boundary values. Then the free unfolded quantized theory
is defined by \eqref{Phi_J_qeq}-\eqref{J_qeq} plus equations for
$\hat{\mathrm{C}}_{ik}^{0}$ 
\begin{equation}
\bigl((N_{i}+1)\mathrm{D}_{i}-(e\partial\bar{\partial})_{i}+(ey\bar{y})_{i}(m^{2}-\frac{\partial}{\partial\tau_{i}})\bigr)\hat{\mathrm{C}}_{ik}^{0}Z=0,\label{c_free_eq_1}
\end{equation}
\begin{equation}
\bigl((N_{k}+1)\mathrm{D}_{k}-(e\partial\bar{\partial})_{k}+(ey\bar{y})_{k}(m^{2}-\frac{\partial}{\partial\tau_{k}})\bigr)\hat{\mathrm{C}}_{ik}^{0}Z=0,\label{c_free_eq_2}
\end{equation}
with the boundary condition in coinciding space-time points $x_{i}=x_{k}$
\begin{equation}
\hat{\mathrm{C}}_{ik}^{0}|_{x_{i}=x_{k}}=i\hbar K_{\tau_{i}+\tau_{k}}^{m}(y_{i}\bar{y}_{i};y_{k}\bar{y}_{k}).\label{c_free_cauchy}
\end{equation}
An unfolded operator algebra is 
\begin{equation}
[\hat{\Phi}_{i},\hat{\Phi}_{k}]=[\hat{\mathrm{J}}_{i},\hat{\mathrm{J}}_{k}]=0,\quad[\hat{\Phi}_{i},\hat{\mathrm{J}}_{k}]=\hat{\mathrm{C}}_{ik}^{0},\label{algebra_free}
\end{equation}
that obviously obeys Jacobi identities.

Let us summarize the results of this Subsection. The unfolded quantum
theory of the free scalar field is determined by the set of unfolded
equations \eqref{Phi_J_qeq}-\eqref{J_qeq}, \eqref{c_free_eq_1}-\eqref{c_free_eq_2},
with the coordinate-independent boundary condition \eqref{c_free_cauchy}
for coinciding space-time points, and by the unfolded operator algebra
\eqref{algebra_free}. In Cartesian coordinates, the commutator takes
a simple form \eqref{F_J_comm_free} and $W$ can be found to have
a usual form \eqref{W_usual}, while an unfolded propagator is \eqref{unf_free_prop},
with a space-time singularity smeared by the heat kernel. Now we turn
to the more complicated case of an interacting theory.

\subsection{Self-interacting quantum scalar\label{SUB_INT_QSCALAR}}

To quantize a nonlinear system, we start with the classical unfolded
off-shell equation \eqref{F_eq} and reformulate it as 
\begin{equation}
(N+1)\mathrm{D}\Phi-e\partial\bar{\partial}\Phi+ey\bar{y}(m^{2}\Phi+g\mathrm{U}'(\Phi)-\mathrm{J})=0\label{F_int_class_eq}
\end{equation}
\begin{equation}
(N+1)\mathrm{D}\mathrm{J}-e\partial\bar{\partial}\mathrm{J}+ey\bar{y}(m^{2}\mathrm{J}-\dot{\mathrm{J}})=0.\label{J_int_class_eq}
\end{equation}
Note that the equation for $\mathrm{J}$ is the free one \eqref{J_eq},
which does not spoil the consistency of \eqref{F_int_class_eq}-\eqref{J_int_class_eq},
but implies that the relation between $\dot{\Phi}$ and $\mathrm{J}$
is much more complicated than in the free case.

We want to promote the system \eqref{F_int_class_eq}-\eqref{J_int_class_eq}
to the operator equations 
\begin{equation}
[(N+1)\mathrm{D}\hat{\Phi}-e\partial\bar{\partial}\hat{\Phi}+ey\bar{y}(m^{2}\hat{\Phi}+g\mathrm{U}'(\hat{\Phi})-\mathrm{\hat{J}})]Z=0,\label{F_int_qeq}
\end{equation}
\begin{equation}
[(N+1)\mathrm{D}-e\partial\bar{\partial}+ey\bar{y}(m^{2}-\frac{\partial}{\partial\tau})]\mathrm{\hat{J}}Z=0.\label{J_int_qeq}
\end{equation}
Assuming \eqref{F_F_comm_J_J_comm} again, the self-consistency of
\eqref{F_int_qeq}-\eqref{J_int_qeq} requires (in Cartesian coordinates)
\begin{equation}
\bigl(\square_{i}+m^{2}+g\mathrm{U}''(\hat{\Phi}_{i})\bigr)[\hat{\Phi}_{i},\hat{\mathrm{J}}_{k}]=\bigl(\square_{k}+m^{2}+g\mathrm{U}''(\hat{\Phi}_{k})\bigr)[\hat{\Phi}_{k},\hat{\mathrm{J}}_{i}],\label{int_cons_1}
\end{equation}
\begin{equation}
\bigl(\square_{i}+m^{2}+g\mathrm{U}''(\hat{\Phi}_{i})\bigr)\bigl(\square_{k}+m^{2}-\frac{\partial}{\partial\tau_{k}}\bigr)[\hat{\Phi}_{i},\hat{\mathrm{J}}_{k}]=0.\label{int_cons_2}
\end{equation}
One can write down a particular solution to \eqref{int_cons_1}-\eqref{int_cons_2}
as 
\begin{equation}
[\hat{\Phi}_{i},\hat{\mathrm{J}}_{k}]=\frac{i\hbar}{(\square_{i}+m^{2}+g\mathrm{U}''(\hat{\Phi}_{i}))}\int d^{4}zK_{\tau_{i}}^{m}(x_{i}+y\bar{y}_{i};z)\bigl(\square_{z}+m^{2}+g\mathrm{U}''(\hat{\Phi}(0,0|z))\bigr)K_{\tau_{k}}^{m}(z;x_{k}+y\bar{y}_{k}).\label{F_J_comm_int}
\end{equation}
\eqref{int_cons_2} is solved due to \eqref{diff_eq_K}, while \eqref{int_cons_1}
is satisfied because the integral in \eqref{F_J_comm_int} is symmetric
under $i\leftrightarrow k$. Moreover, \eqref{F_J_comm_int} still
has the correct form \eqref{primary_comm} in the primary ($\tau=0$,
$Y=0$)-sector. Note that the commutator \eqref{F_J_comm_int} is
not symmetric with respect to an exchange of the arguments of operators,
$[\hat{\Phi}_{i},\hat{\mathrm{J}}_{k}]\neq[\hat{\Phi}_{k},\hat{\mathrm{J}}_{i}]$,
in contrast to the standard QFT \eqref{primary_comm} and unfolded
free theory \eqref{F_J_comm_free} cases.

The commutator \eqref{F_J_comm_int} has a quite complicated structure,
being non-local and field-dependent. To the first order in $g$ it
is 
\begin{equation}
[\hat{\Phi}_{i},\hat{\mathrm{J}}_{k}]=\hat{\mathrm{C}}_{ik}^{0}-\frac{g}{(\square_{i}+m^{2})}\mathrm{U}''(\hat{\Phi}_{i})\hat{\mathrm{C}}_{ik}^{0}-\frac{ig}{\hbar(\square_{i}+m^{2})}\int d^{4}x_{u}\hat{\mathrm{C}}_{iu}^{0}\mathrm{U}''(\hat{\Phi}_{u})\hat{\mathrm{C}}_{uk}^{0}|_{Y_{u},\tau_{u}=0}+O(g^{2}),\label{corr_int_perturb}
\end{equation}
where $\hat{\mathrm{C}}_{ik}^{0}$ is the commutator of the free theory
\eqref{F_J_comm_free}.

It is possible to formulate the quantized nonlinear theory in a coordinate-independent
way as well. One introduces a two-point unfolded operator field 
\begin{equation}
\hat{\mathrm{C}}_{ik}=[\hat{\Phi}_{i},\hat{\mathrm{J}}_{k}],
\end{equation}
subjected to 
\begin{equation}
[(N_{i}+1)\mathrm{D}_{i}-(e\partial\bar{\partial})_{i}+(ey\bar{y})_{i}(m^{2}+g\mathrm{U}''(\hat{\Phi}_{i}))]\hat{\mathrm{C}}_{ik}Z=(ey\bar{y})_{i}\hat{f}_{ik}Z,\label{c_int_eq_1}
\end{equation}
\begin{equation}
[(N_{k}+1)\mathrm{D}_{k}-(e\partial\bar{\partial})_{k}+(ey\bar{y})_{k}(m^{2}-\frac{\partial}{\partial\tau_{k}})]\hat{\mathrm{C}}_{ik}Z=0,\label{c_int_eq_2}
\end{equation}
where $\hat{f}_{ik}=\hat{f}(\tau_{i},x_{i}+y\bar{y}_{i};\tau_{k},x_{k}+y\bar{y}_{k})$
is an arbitrary two-point operator, symmetric in its arguments $\hat{f}_{ik}=\hat{f}_{ki}$.
Specifying $\hat{f}$ determines $\hat{\mathrm{C}}$. In our case,
in Cartesian coordinates, where $\hat{\mathrm{C}}$ is \eqref{F_J_comm_int},
we have 
\begin{equation}
\hat{f}_{ik}=i\hbar\int d^{4}zK_{\tau_{i}}^{m}(x_{i}+y\bar{y}_{i};z)\bigl(\square_{z}+m^{2}+g\mathrm{U}''(\hat{\Phi}(0,0|z))\bigr)K_{\tau_{k}}^{m}(z;x_{k}+y\bar{y}_{k}).\label{f_ik}
\end{equation}
But instead of trying to find a coordinate-independent description
of \eqref{f_ik}, we impose coordinate-independent boundary conditions
for $\mathrm{\hat{C}}$, leaving $\hat{f}$ undetermined. They are
\begin{equation}
\hat{\mathrm{C}}_{ik}|_{\tau_{i}=0;x_{i}=x_{k}}=i\hbar K_{\tau_{k}}^{m}(y\bar{y}_{i};y\bar{y}_{k}),\label{c_int_cauchy}
\end{equation}
plus a free-limit constraint, that $\hat{\mathrm{C}}_{ik}(g=0)$ turns
to the commutator of the free theory, determined by \eqref{c_free_eq_1}-\eqref{c_free_cauchy}.

An unfolded operator algebra looks like in the free case, 
\begin{equation}
[\hat{\Phi}_{i},\hat{\Phi}_{k}]=[\hat{\mathrm{J}}_{i},\hat{\mathrm{J}}_{k}]=0,\quad[\hat{\Phi}_{i},\hat{\mathrm{J}}_{k}]=\hat{\mathrm{C}}_{ik},
\end{equation}
but the check of Jacobi identities is more complicated now. Namely,
the identity for $[[\hat{\Phi},\hat{\mathrm{J}}],\hat{\mathrm{J}}]$
is not obvious anymore, because now $\hat{\mathrm{C}}$ is $\hat{\Phi}$-dependent.
One should use the standard relations for the commutators in an associative
algebra 
\begin{equation}
[AB,C]=A[B,C]+[A,C]B,\quad[A^{-1},B]=-A^{-1}[A,B]A^{-1},
\end{equation}
in order to find 
\begin{eqnarray}
[\hat{\mathrm{C}}_{ij},\hat{\mathrm{J}}_{k}] & = & \frac{(i\hbar)^{2}g}{(\square_{i}+m^{2}+g\mathrm{U}''(\hat{\Phi}_{i}))}\Bigl(-\mathrm{U}'''(\hat{\Phi}_{i})\hat{\mathrm{C}}_{ij}\hat{\mathrm{C}}_{ik}+\nonumber \\
 &  & +\int d^{4}z\mathrm{U}'''(\hat{\Phi}(0,0|z))K_{\tau_{i}}^{m}(x_{i}+y\bar{y}_{i};z)K_{\tau_{j}}^{m}(x_{j}+y\bar{y}_{j};z)K_{\tau_{k}}^{m}(x_{k}+y\bar{y}_{k};z)\Bigr)\label{C_J_comm}
\end{eqnarray}
which is manifestly symmetric under $j\leftrightarrow k$ and hence
guarantees the Jacobi identity for $[[\hat{\Phi},\hat{\mathrm{J}}],\hat{\mathrm{J}}]$.

Thus, the unfolded self-interacting scalar can be quantized along
the same lines as the free one, but the commutator of $\hat{\Phi}$
and $\hat{\mathrm{J}}$ in this case becomes much more complicated,
representing a $\hat{\Phi}$-dependent nonlocal expression. This makes
the problem of solving for $Z$ too difficult. Therefore, it looks
reasonable to use the system we built, in order to formulate equations
directly for unfolded correlation functions. This is done in the next
Section.

\section{Unfolded correlation functions\label{SEC_CORREL}}

Starting from the functional Schwinger--Dyson equation \eqref{SD_eq},
one can deduce standard Schwinger --Dyson equations for correlation
functions. Let us rewrite \eqref{SD_eq} in the operator form 
\begin{equation}
\frac{\delta S}{\delta\phi_{i}}[\hat{\phi}]Z=\hat{j}_{i}Z\label{SD_eq-1}
\end{equation}
(here $\phi_{i}\equiv\phi(x_{i})$) with \eqref{primary_comm} imposed.
Acting on \eqref{SD_eq-1} with field operators $\hat{\phi}_{a_{1}}$,
$\hat{\phi}_{a_{2}}$, ... $\hat{\phi}_{a_{n}}$ and putting $\hat{j}=0$
at the end, we find 
\begin{equation}
\hat{\phi}_{a_{1}}\hat{\phi}_{a_{2}}...\hat{\phi}_{a_{n}}\frac{\delta S}{\delta\phi_{i}}[\hat{\phi}]Z|_{\hat{j}=0}=\sum_{k=1}^{n}\hat{\phi}_{a_{1}}\hat{\phi}_{a_{2}}...[\hat{\phi}_{a_{k}},\hat{j}_{i}]...\hat{\phi}_{a_{n}}Z|_{\hat{j}=0}.\label{SD_eq-1-1}
\end{equation}
Considering that the action of field operators on $Z$ at zero sources
produces corresponding correlation functions, 
\begin{equation}
\hat{\phi}_{a_{1}}\hat{\phi}_{a_{2}}...\hat{\phi}_{a_{n}}Z|_{\hat{j}=0}=\bigl\langle\phi_{a_{1}}\phi_{a_{2}}...\phi_{a_{n}}\bigr\rangle,\label{correl_def}
\end{equation}
one makes use of \eqref{primary_comm} and recovers from \eqref{SD_eq-1-1}
the Schwinger--Dyson equation for correlation functions, 
\begin{equation}
\bigl\langle\frac{\delta S}{\delta\phi_{i}}\phi_{a_{1}}\phi_{a_{2}}...\phi_{a_{n}}\bigr\rangle=i\hbar\sum_{k=1}^{n}\bigl\langle\phi_{a_{1}}\phi_{a_{2}}...\delta^{4}(x_{i}-x_{a_{k}})...\phi_{a_{n}}\bigr\rangle.\label{SD_eq-1-1-1}
\end{equation}

Our goal is to provide an analogue of \eqref{SD_eq-1-1-1} for unfolded
fields, so one can perturbatively solve for unfolded correlators.
We naturally define an unfolded $n$-point correlation function to
be 
\begin{equation}
\bigl\langle\Phi_{a_{1}}\Phi_{a_{2}}...\Phi_{a_{n}}\bigr\rangle=\hat{\Phi}_{a_{1}}\hat{\Phi}_{a_{2}}...\hat{\Phi}_{a_{n}}Z|_{\mathrm{J}=0},
\end{equation}
which in the primary ($\tau=0$, $Y=0$)-sector coincides with \eqref{correl_def}.
Acting with $\hat{\Phi}_{a_{1}}\hat{\Phi}_{a_{2}}...\hat{\Phi}_{a_{n}}$
on \eqref{F_int_qeq}, which is an unfolded substitute for \eqref{SD_eq-1},
we have 
\begin{eqnarray}
 &  & \left((N_{i}+1)\mathrm{D}_{i}-(e\partial\bar{\partial})_{i}+(ey\bar{y})_{i}m^{2}\right)\bigl\langle\Phi_{i}\Phi_{a_{1}}\Phi_{a_{2}}...\Phi_{a_{n}}\bigr\rangle+\nonumber \\
 &  & +(ey\bar{y})_{i}\left(g\bigl\langle\mathcal{\mathrm{U}}'(\Phi_{i})\Phi_{a_{1}}\Phi_{a_{2}}...\Phi_{a_{n}}\bigr\rangle+\sum_{k=1}^{n}\bigl\langle\Phi_{a_{1}}\Phi_{a_{2}}...\mathrm{C}_{a_{k},i}...\Phi_{a_{n}}\bigr\rangle\right)=0.\label{unf_SD_corr}
\end{eqnarray}
with $\hat{\mathrm{C}}_{a_{k},i}\equiv[\hat{\Phi}_{a_{k}},\hat{\mathrm{J}}_{i}]$
from Subsection \ref{SUB_INT_QSCALAR}. The system \eqref{unf_SD_corr}
provides an unfolded form of a chain of Schwinger--Dyson equations
and allows one to iteratively calculate unfolded correlation functions.

For the free theory with $\mathrm{U}'=0$, \eqref{unf_SD_corr} in
Cartesian coordinates reads as 
\begin{eqnarray}
 &  & \left((N_{i}+1)\mathrm{d}_{i}-(e\partial\bar{\partial})_{i}+(ey\bar{y})_{i}m^{2}\right)\bigl\langle\Phi_{i}\Phi_{a_{1}}\Phi_{a_{2}}...\Phi_{a_{n}}\bigr\rangle^{0}+\nonumber \\
 &  & +i\hbar(ey\bar{y})_{i}\sum_{k=1}^{n}\bigl\langle\Phi_{a_{1}}\Phi_{a_{2}}...K_{\tau_{i}+\tau_{a_{k}}}^{m}(x_{i}+y\bar{y}_{i};x_{a_{k}}+y\bar{y}_{a_{k}})...\Phi_{a_{n}}\bigr\rangle^{0}=0.\label{SD_eq_free}
\end{eqnarray}
As expected, for the unfolded fields the contact terms get smoothed,
having the heat kernel in place of the delta-function. In the nonlinear
problem, the contact terms get additional field dressing, due to $\hat{\Phi}$-dependence
of $\mathrm{\hat{C}}$.

Let us use \eqref{unf_SD_corr} to calculate a first-order perturbative
correction to the unfolded propagator. To this end we expand a full
propagator in a coupling constant as 
\begin{equation}
\bigl\langle\Phi_{i}\Phi_{k}\bigr\rangle=\bigl\langle\Phi_{i}\Phi_{k}\bigr\rangle^{0}+g\bigl\langle\Phi_{i}\Phi_{k}\bigr\rangle^{g}+O(g^{2}).
\end{equation}

An exact equation for the full propagator is 
\begin{equation}
\left((N_{i}+1)\mathrm{D}_{i}-(e\partial\bar{\partial})_{i}+(ey\bar{y})_{i}m^{2}\right)\bigl\langle\Phi_{i}\Phi_{k}\bigr\rangle+g(ey\bar{y})_{i}\bigl\langle\mathcal{\mathrm{U}}'(\Phi_{i})\Phi_{k}\bigr\rangle+i\hbar(ey\bar{y})_{i}\bigl\langle\mathrm{C}_{ki}\bigr\rangle=0.
\end{equation}
In zeroth order in $g$ and in Cartesian coordinates this yields 
\begin{equation}
\left((N_{i}+1)\mathrm{d}_{i}-(e\partial\bar{\partial})_{i}+(ey\bar{y})_{i}m^{2}\right)\bigl\langle\Phi_{i}\Phi_{k}\bigr\rangle^{0}+i\hbar(ey\bar{y})_{i}K_{\tau_{i}+\tau_{k}}^{m}(x_{i}+y\bar{y}_{i};x_{k}+y\bar{y}_{k})=0,
\end{equation}
whose solution, of course, coincides with the free propagator \eqref{unf_free_prop}.

Then for the first-order correction we have, using \eqref{corr_int_perturb},
\begin{eqnarray}
 &  & \left((N_{i}+1)\mathrm{d}_{i}-(e\partial\bar{\partial})_{i}+(ey\bar{y})_{i}m^{2}\right)\bigl\langle\Phi_{i}\Phi_{k}\bigr\rangle^{g}+(ey\bar{y})_{i}\bigl\langle\mathcal{\mathrm{U}}'(\Phi_{i})\Phi_{k}\bigr\rangle^{0}+\nonumber \\
 &  & +(ey\bar{y})_{i}\frac{1}{\square_{k}+m^{2}}\left(\bigl\langle\mathcal{\mathrm{U}}''(\Phi_{k})\bigr\rangle^{0}\mathrm{C}_{k,i}^{0}+\frac{i}{\hbar}\int d^{4}x_{u}\mathrm{C}_{k,u}^{0}\bigl\langle\mathcal{\mathrm{U}}''(\Phi_{u})\bigr\rangle^{0}\mathrm{C}_{u,i}^{0}|_{Y_{u},\tau_{u}=0}\right)=0.
\end{eqnarray}
The free equation \eqref{SD_eq_free} allows us to find 
\begin{equation}
\bigl\langle\mathcal{\mathrm{U}}'(\Phi_{i})\Phi_{k}\bigr\rangle^{0}=\frac{1}{\square_{k}+m^{2}}\mathrm{C}_{i,k}^{0}\bigl\langle\mathcal{\mathrm{U}}''(\Phi_{i})\bigr\rangle^{0}.
\end{equation}
For a translation-invariant ground state, correlation functions of
$\Phi$ taken in coinciding points $(Y,\tau|x)$ can only depend on
$\tau$, therefore 
\begin{equation}
\bigl\langle\mathcal{\mathrm{U}}''(\Phi_{i})\bigr\rangle^{0}=:\Lambda^{2}(\tau_{i}),\label{Lambda_def}
\end{equation}
and finally we arrive at 
\begin{equation}
\bigl\langle\Phi_{i}\Phi_{k}\bigr\rangle^{g}=-\frac{i\hbar\left(\Lambda^{2}(\tau_{i})+\Lambda^{2}(\tau_{k})-\Lambda^{2}(0)\right)}{(\square_{i}+m^{2})(\square_{k}+m^{2})}e^{m^{2}(\tau_{i}+\tau_{k})}K_{\tau_{i}+\tau_{k}}(x_{i}+y\bar{y}_{i};x_{k}+y\bar{y}_{k}).\label{unf_propag_corr}
\end{equation}
The first-order correction to the unfolded propagator \eqref{unf_propag_corr}
generalizes the standard QFT result, which arises upon sending all
$\tau$ and $Y$ to zero. The latter usually diverges, containing
a factor $\Lambda^{2}(0)=\bigl\langle\mathcal{\mathrm{U}}''(\phi(x))\bigr\rangle^{0}$,
and therefore requires regularization. This factor is presented in
the unfolded expression \eqref{unf_propag_corr} as well, but now
in a special additive combination $\left(\Lambda^{2}(\tau_{i})+\Lambda^{2}(\tau_{k})-\Lambda^{2}(0)\right)$.
This may have interesting consequences for the problem of renormalizations
of the unfolded QFT: we see that, in effect, $\Lambda^{2}(\tau)$
defined by \eqref{Lambda_def} serves as a built-in regularizer of
the unfolded theory, smearing the divergent self-energy $\Lambda^{2}(0)$
with the proper time $\tau$. And although the unfolded expression
\eqref{unf_propag_corr} still diverges, now one sees that the problem
of renormalization amounts to the redefinition of $\Lambda^{2}(\tau)$.
The systematic development of perturbative calculation and renormalization
methods for the unfolded QFT requires a separate thorough consideration,
and we leave this for future work.

\section{Semiclassical quantization and unfolded effective equations\label{SEC_LOOP}}

Starting from the unfolded system for the partition function $Z$,
formulated in Section \ref{SEC_SD}, it is possible to construct a
different unfolded realization of the QFT, which involves a quantum
effective action. It allows one to calculate iteratively (in powers
of $\hbar$) unfolded one-particle irreducible vertex functions. They
generalize 1PI vertex functions of standard QFT in the same way as
unfolded correlators of Section \ref{SEC_CORREL} generalize the standard
ones.

Let us begin with a short reminder of the standard construction. Quantum
effective action $\Gamma[\bar{\phi}]$ is a generating functional
for 1PI vertex functions. It is defined as the Legendre transform
of a generator of connected correlators $W$ \eqref{W_def} 
\begin{equation}
\Gamma[\bar{\phi}]=\int d^{4}x\bar{\phi}j+W[j],\label{Gamma_Legendre}
\end{equation}
where a mean (or classical) field $\bar{\phi}(x)$ represents an expectation
value of the corresponding quantum field at non-vanishing sources
\begin{equation}
\bar{\phi}(x)=\bigl\langle\phi(x)\bigr\rangle_{J}.
\end{equation}
Expansion of $\Gamma$ in powers of the Planck constant corresponds
to the expansion in loop corrections to the classical action $S$.

One can deduce an equation that determines $\Gamma$ from the functional
Schwinger--Dyson equation \eqref{SD_eq}. To this end, one uses \eqref{W_def},
\eqref{Gamma_Legendre} and 
\begin{equation}
j(x)=\frac{\delta\Gamma}{\delta\bar{\phi}(x)},\label{gamma_var}
\end{equation}
following from \eqref{Gamma_Legendre}, to represent $Z$ as a functional
of the mean field 
\begin{equation}
Z[j(\bar{\phi})]=\exp\left(\frac{i}{\hbar}(1-\int d^{4}x\bar{\phi}(x)\frac{\delta}{\delta\bar{\phi}(x)})\Gamma[\bar{\phi}]\right).\label{Z_mean}
\end{equation}
From \eqref{gamma_var}, one can also express $j$-variational derivative
through $\bar{\phi}$-variational derivative as 
\begin{equation}
\frac{\delta}{\delta j(\bar{\phi})}=\int d^{4}y\left(\frac{\delta^{2}\Gamma}{\delta\bar{\phi}(x)\delta\bar{\phi}(y)}\right)^{-1}\frac{\delta}{\delta\bar{\phi}(y)}.\label{dj_mean}
\end{equation}
Substituting \eqref{gamma_var}, \eqref{Z_mean} and \eqref{dj_mean}
to the functional Schwinger--Dyson equation \eqref{SD_eq}, one finds
the following equation for $\Gamma$ 
\begin{equation}
\frac{\delta S}{\delta\phi}[\bar{\phi}(x)+i\hbar\int d^{4}y(\frac{\delta^{2}\Gamma}{\delta\bar{\phi}(x)\delta\bar{\phi}(y)})^{-1}\frac{\delta}{\delta\bar{\phi}(y)}]=\frac{\delta\Gamma}{\delta\bar{\phi}(x)},\label{Gamma_eq}
\end{equation}
where $\bar{\phi}$-derivative acts to the right. This equation determines
the effective action up to a field-independent contribution $\Gamma[0]$.

As is seen from \eqref{Gamma_eq}, this equation can be easily obtained
from the classical e.o.m. coupled to an external source \eqref{class_eq}.
One should simply replace 
\begin{equation}
\phi(x)\rightarrow\bar{\phi}(x)+i\hbar\frac{\delta}{\delta j(x)}\label{mean_quant}
\end{equation}
and substitute \eqref{gamma_var}. Then one arrives at \eqref{Gamma_eq}.
Note, however, that while the shift \eqref{mean_quant} can always
be performed for arbitrary e.o.m., the substitute \eqref{gamma_var}
requires from the e.o.m. to be Lagrangian. For the models of self-interacting
scalar we study in the paper, this is always true, but in general
one should check this for the consistency of quantization. A systematic
procedure to analyze Lagrangian properties of the unfolded systems
is presented in \citep{unf_Lagr}.

For our needs it is more convenient to reformulate \eqref{Gamma_eq}
in the form without variations, like was done in Subsection \ref{SUB_FUNCT_SD}.
To this end, we define a classical Poisson bracket 
\begin{equation}
\{j(x_{1}),\phi(x_{2})\}=\delta^{4}(x_{1}-x_{2}),\quad\{\phi(x_{1}),\phi(x_{2})\}=\{j(x_{1}),j(x_{2})\}=0,\label{Poiss_bracket}
\end{equation}
which obviously satisfies Jacobi identity. Then the equation for the
effective action can be reformulated in the form 
\begin{equation}
\frac{\delta S}{\delta\phi}[\bar{\phi}(x)+i\hbar\int d^{4}yG_{xy}\{j(y),\bullet\}]=\{j(x),\Gamma\},\label{Gamma_eq-1}
\end{equation}
where $\{j(y),\bullet\}$ acts to the right and a propagator $G_{xy}$
is an inverse to $\Gamma_{xy}$ 
\begin{equation}
\int d^{4}y\Gamma_{xy}G_{yz}=\delta^{4}(x-z),\quad\Gamma_{xy}[\bar{\phi}]=\{j(y),\{j(x),\Gamma\}\}.\label{Green_def}
\end{equation}
Note that the symmetricity of $\Gamma_{xy}$ is ensured by the Jacobi
identity for the Poisson bracket \eqref{Poiss_bracket}. The same
is true for higher vertex functions $\Gamma_{x_{1}x_{2}...x_{n}}$,
which result from the successive application of $\{j(x_{i}),\bullet\}$
to $\Gamma$.

Separating a classical part $G_{xy}^{0}$ of the full propagator,
\begin{equation}
G_{xy}=G_{xy}^{0}+O(\hbar),\quad\int d^{4}yS_{xy}G_{yz}^{0}=\delta^{4}(x-z),
\end{equation}
which is determined by the classical action $S[\phi]$, from \eqref{Green_def}
one can deduce a relation 
\begin{equation}
G_{xy}=G_{xy}^{0}-\int d^{4}z_{1}d^{4}z_{2}G_{xz_{1}}^{0}(\Gamma_{z_{1}z_{2}}-S_{z_{1}z_{2}})G_{z_{2}y},\label{G_eq_qft}
\end{equation}
which allows for an iterative recovering of the $\hbar$-corrections
to $G_{xy}^{0}$.

Our goal is to formulate a system of effective unfolded equations,
which determines $\Gamma$ and allows one to calculate unfolded vertex
functions.

We start with a classical unfolded system 
\begin{equation}
(N+1)\mathrm{D}\Phi-e\partial\bar{\partial}\Phi+ey\bar{y}(m^{2}\Phi+g\mathrm{U}'(\Phi)-\mathrm{J})=0,\label{class_eq_1}
\end{equation}
\begin{equation}
(N+1)\mathrm{D}\mathrm{J}-e\partial\bar{\partial}\mathrm{J}+ey\bar{y}(m^{2}\mathrm{J}-\dot{\mathrm{J}})=0,\label{class_eq_2}
\end{equation}
and, introducing a notation 
\begin{equation}
\hat{\mathrm{A}}_{i}\equiv\hat{\mathrm{A}}(Y_{i},\tau_{i}|x_{i}):=\square_{i}+m^{2}+g\mathrm{U}''(\Phi_{i}),
\end{equation}
define an unfolded classical Poisson bracket, 
\begin{equation}
\pi_{ik}^{0}\equiv\{\mathrm{J}_{i},\Phi_{k}\}^{0}=\hat{\mathrm{A}}_{k}^{-1}\int d^{4}zK_{\tau_{k}}^{m}(x_{k}+y_{k}\bar{y}_{k};z)\hat{\mathrm{A}}(0,0|z)K_{\tau_{i}}^{m}(z;x_{i}+y_{i}\bar{y}_{i}),\label{F_J_PB}
\end{equation}
\begin{equation}
\{\Phi_{i},\Phi_{k}\}=\{\mathrm{J}_{i},\mathrm{J}_{k}\}=0,\label{F_J_PB-2}
\end{equation}
which is nothing but an image of the commutator \eqref{F_J_comm_int}
of the corresponding quantum operators. We work here in Cartesian
coordinates for simplicity, but one can write down a coordinate-independent
formulation of the Poisson bracket algebra, in a straightforward analogy
with Subsection \ref{SUB_INT_QSCALAR}.

We quantize the classical system \eqref{class_eq_1}-\eqref{class_eq_2}
with the Poisson bracket \eqref{F_J_PB}-\eqref{F_J_PB-2} by replacing
in \eqref{class_eq_1} 
\begin{equation}
\mathrm{J}_{i}\rightarrow\Gamma_{i}\equiv\{\mathrm{J}_{i},\Gamma\},\quad\mathrm{U}'(\Phi_{i})\rightarrow\mathrm{U'_{eff}}(\Phi_{i})\equiv\mathrm{U'}(\Phi_{i}+i\hbar G_{ik}\overset{k}{\circ}\{\mathrm{J}_{k},\bullet\}),\label{wkb_quant}
\end{equation}
where $\Gamma[\Phi]$ is an unfolded quantum effective action and
an unfolded propagator $G_{ik}=G_{ki}$ is related to its inverse
as 
\begin{equation}
\Gamma_{ij}\overset{j}{\circ}G_{jk}=\pi_{ik},\quad G_{ij}\overset{j}{\circ}\Gamma_{jk}=\pi_{ki},\quad\Gamma_{ik}\equiv\{\mathrm{J}_{i},\Gamma_{k}\},\label{G_Ginv_K}
\end{equation}
where $\pi_{ik}$ is a quantum generalization of the classical bracket
$\pi_{ik}^{0}$ \eqref{F_J_PB} and is discussed below. A contraction
operation $\circ$ in \eqref{wkb_quant} and \eqref{G_Ginv_K} is
defined as 
\begin{equation}
F_{i}\overset{i}{\circ}G_{i}:=\int d^{4}x_{i}F(\tau_{i}=0;x_{i})G(\tau_{i}=0;x_{i}),
\end{equation}
that corresponds to the integration of primary components of unfolded
scalar fields $F$ and $G$ (note that $Y$-dependence vanishes automatically,
because it has the form $x+y\bar{y}$ for the functions in question).

Finally, unfolded quantum equations for the mean field and the source
are 
\begin{equation}
(N_{i}+1)\mathrm{D}_{i}\Phi_{i}-(e\partial\bar{\partial})_{i}\Phi_{i}+(ey\bar{y})_{i}\left(m^{2}\Phi_{i}+g\mathrm{U'_{eff}}(\Phi_{i})-\Gamma_{i}\right)=0,\label{eff_eq_1}
\end{equation}
\begin{equation}
(N_{i}+1)\mathrm{D}_{i}\mathrm{J}_{i}-(e\partial\bar{\partial})_{i}\mathrm{J}_{i}+(ey\bar{y})_{i}(m^{2}\mathrm{J}_{i}-\dot{\mathrm{J}}_{i})=0.\label{eff_eq_2}
\end{equation}
These must be supported with unfolded equations determining $\pi_{ik}=\{\mathrm{J}_{i},\Phi_{k}\}$,
which in general is different from the classical bracket \eqref{F_J_PB}.
To find them, one takes the bracket of \eqref{eff_eq_1} with $\mathrm{J}$,
producing an equation for the second argument of $\pi_{ik}$ 
\begin{equation}
((N_{k}+1)\mathrm{D}_{k}-(e\partial\bar{\partial})_{k}+m^{2}(ey\bar{y})_{k})\pi_{ik}+(ey\bar{y})_{k}\left(g\{\mathrm{J}_{i},\mathrm{U'_{eff}}(\Phi_{k})\}-\Gamma_{ik}\right)=0,\label{eff_eq_3}
\end{equation}
and takes the bracket of \eqref{eff_eq_2} with $\Phi$, producing
an equation for the first argument 
\begin{equation}
((N_{i}+1)\mathrm{D}_{i}-(e\partial\bar{\partial})_{i}+(ey\bar{y})_{i}(m^{2}-\frac{\partial}{\partial\tau_{i}}))\pi_{ik}=0.\label{eff_eq_4}
\end{equation}

An analogue of \eqref{G_eq_qft} for the unfolded system is 
\begin{equation}
\pi_{ia}^{0}\overset{i}{\circ}G_{ib}=G_{ai}^{0}\overset{i}{\circ}\pi_{ib}-gG_{ai}^{0}\overset{i}{\circ}\{\mathrm{J}_{i},\mathrm{U'_{eff}}(\Phi_{j})\}\overset{j}{\circ}G_{jb}+gG_{ai}^{0}\mathrm{U''}(\Phi_{i})\overset{i}{\circ}G_{ib},
\end{equation}
\begin{equation}
G_{ai}\overset{i}{\circ}\pi_{ib}^{0}=\pi_{ia}\overset{i}{\circ}G_{ib}^{0}-gG_{ai}\overset{i}{\circ}\{\mathrm{J}_{i},\mathrm{U'_{eff}}(\Phi_{j})\}\overset{j}{\circ}G_{jb}^{0}+gG_{ai}\mathrm{U''}(\Phi_{i})\overset{i}{\circ}G_{ib}^{0},
\end{equation}
and a classical propagator can be checked to be 
\begin{equation}
G_{ab}^{0}=\hat{\mathrm{A}}_{a}^{-1}\pi_{ab}^{0}.
\end{equation}

The system of unfolded quantum effective equations \eqref{eff_eq_1}-\eqref{eff_eq_4}
requires the definition of concrete $\pi_{ik}$, that obeys \eqref{eff_eq_3}-\eqref{eff_eq_4}
and Jacobi identities and leads to symmetric $\Gamma_{ik}$. A special
feature of \eqref{eff_eq_3} is that the bracket $\pi_{ik}$ is determined
by the effective potential $\mathrm{U'_{eff}}$ \eqref{wkb_quant},
which in its turn is determined by $\pi_{ik}$. The way out is to
work within the semiclassical expansion, restoring $\pi_{ik}$ order
by order in $\hbar$, with \eqref{F_J_PB} being the zeroth term.

$\hbar$-term in $\mathrm{U'_{eff}}(\Phi)$ in \eqref{wkb_quant}
results in the appearance of the higher unfolded vertex functions
\begin{equation}
\Gamma_{i_{1}...i_{n}}=\{\mathrm{J}_{i_{n}}...\{\mathrm{J}_{i_{2}}\{\mathrm{J}_{i_{1}},\Gamma\}\}\}
\end{equation}
(whose symmetricity is ensured by the Jacobi identity for $\{\bullet,\bullet\}$),
for which one needs new unfolded equations, in order to have a closed
system. These equations can be obtained by the successive application
of $\{\mathrm{J},\bullet\}$ to \eqref{eff_eq_1}. In general, vertex
functions of all orders contribute to \eqref{eff_eq_1}, which leads
to an infinite system of entangled unfolded equations for all $\Gamma_{i_{1}...i_{n}}$.
But when one restricts oneself to some specific order in $\hbar$,
the system reduces to a finite number of equations. Thus again, from
the practical point of view, the unfolded effective system \eqref{eff_eq_1}-\eqref{eff_eq_4}
should be analyzed within the frame of the semiclassical expansion.
A similar idea of evaluating loop corrections via formulating Schwinger--Dyson
equations in terms of an infinite set of descendant fields has been
put forward in \citep{Lee_q_off_shell}.

To illustrate how the analysis of the system \eqref{eff_eq_1}-\eqref{eff_eq_2}
goes, let us use it to calculate a one-loop correction to the inverse
unfolded propagator $\Gamma_{ik}$.

We choose the following particular (and implicit) solution to \eqref{eff_eq_3}-\eqref{eff_eq_4}
in Cartesian coordinates 
\begin{eqnarray}
 & \pi_{ik}= & \pi_{ik}^{0}+g\hat{\mathrm{A}}_{k}^{-1}(\{\mathrm{J}_{k},\mathrm{U'_{eff}}(\Phi_{i})\}-\mathrm{U''}(\Phi_{i})\pi_{ki})-\nonumber \\
 &  & -\frac{g}{2}\hat{\mathrm{A}}_{k}^{-1}\int d^{4}z\int d^{4}z'K_{\tau_{i}}^{m}(x_{i}+y\bar{y}_{i};z)\Bigl(\{j(z'),\mathrm{U'_{eff}}(\phi(z))\}+\{j(z),\mathrm{U'_{eff}}(\phi(z'))\}-\nonumber \\
 &  & -2\mathrm{U''}(\phi(z))\delta(z-z'))\Bigr)K_{\tau_{k}}^{m}(z';x_{k}+y\bar{y}_{k}).\label{pi_quant}
\end{eqnarray}
Here the bracket $\pi_{ik}$ is expressed in terms of itself, so one
has to perform a semiclassical expansion. We expand up to a linear
order in $\hbar$ 
\begin{equation}
\Gamma_{ik}[\Phi]=S_{ik}+\hbar\Gamma_{ik}^{\hbar}+O(\hbar^{2}),\quad\pi_{ik}=\pi_{ik}^{0}+\hbar\pi_{ik}^{\hbar}+O(\hbar^{2}),\quad\mathrm{U'_{eff}}(\Phi_{i})=\mathrm{U}'(\Phi_{i})+\hbar\mathrm{U'_{\hbar}}(\Phi_{i})+O(\hbar^{2}),
\end{equation}
and for the variation of a one-loop effective potential one finds
\begin{equation}
\mathrm{U'_{\hbar}}(\Phi_{i})=\frac{i}{2}\mathrm{U}'''(\Phi_{i})G_{ii}^{0}.
\end{equation}
Let us note that here the proper time $\tau$ again plays the role
of the regularizer, so that the propagator in coinciding points 
\begin{equation}
G_{ii}^{0}=\int d^{4}z\left(\hat{\mathrm{A}}_{i}^{-1}K_{\tau_{i}}^{m}(z;x_{i}+y\bar{y}_{i})\right)\hat{\mathrm{A}}(0,0|z)\left(\hat{\mathrm{A}}_{i}^{-1}K_{\tau_{i}}^{m}(z;x_{i}+y\bar{y}_{i})\right)
\end{equation}
is singular only when $\tau_{i}\rightarrow0$.

Using this, from the exact equation \eqref{eff_eq_3} one calculates
a one-loop correction $\Gamma_{ik}^{\hbar}$ 
\begin{eqnarray}
 &  & \Gamma_{ik}^{\hbar}=g\{\mathrm{J}_{i},\mathrm{U'_{\hbar}}(\Phi_{k})\}^{0}-\frac{g}{2}\int d^{4}z\int d^{4}z'K_{\tau_{i}}^{m}(x_{i}+y\bar{y}_{i};z')\{j(z'),\mathrm{U'_{\hbar}}(\phi(z)\}^{0}K_{\tau_{k}}^{m}(z;x_{k}+y\bar{y}_{k})+\left(i\leftrightarrow k\right),\nonumber \\
\label{G_ik}
\end{eqnarray}
with 
\begin{eqnarray}
 &  & \{\mathrm{J}_{i},\mathrm{U'_{\hbar}}(\Phi_{k})\}^{0}=\frac{i\hbar}{2}\mathrm{U^{(IV)}}(\Phi_{k})G_{kk}^{0}\pi_{ik}^{0}-i\hbar g\mathrm{U'''}(\Phi_{k})\int d\xi_{u}\left(\hat{\mathrm{A}}_{k}^{-1}\delta(\xi_{k}-\xi_{u})\right)g\mathrm{U}'''(\Phi_{u})\pi_{iu}^{0}G_{uk}^{0}+\nonumber \\
 &  & +\frac{i\hbar g}{2}\mathrm{U'''}(\Phi_{k})\int d^{4}z\left(\hat{\mathrm{A}}_{k}^{-1}K_{\tau_{k}}^{m}(z;x_{k}+y\bar{y}_{k})\right)^{2}\mathrm{U}'''(\phi(z))K_{\tau_{i}}^{m}(z;x_{i}+y\bar{y}_{i}),
\end{eqnarray}
where $\xi$ stands for the full set of coordinates, 
\begin{equation}
\xi_{k}:=(\tau_{k},Y_{k}|x_{k}).
\end{equation}

Using that both the heat kernel and $\pi_{ik}^{0}$ come down to space-time
delta-functions when $(\tau,Y)\rightarrow0$, one can see that in
this limit \eqref{G_ik} correctly reproduces a standard QFT expression.

\section{$5d$ auxiliary model and $\tau$ as a physical time\label{SEC_HOL}}

Finally, let us briefly address the following issue: in a nutshell,
quantization procedures we performed consisted in taking the equation
\begin{equation}
\square\Phi+m^{2}\Phi+g\mathrm{U}'(\Phi)=\dot{\Phi}\label{hol_Phi_eq}
\end{equation}
and quantizing it by identifying $\dot{\Phi}$ with momentum conjugate
to the field $\Phi$; one may wonder if there is any model which has
$\tau$ as a physical time and somehow leads to the equation \eqref{hol_Phi_eq}.

Such model does exist. In order to construct it, we first notice,
that upon identifying $\dot{\Phi}$ with the momentum, one cannot
treat \eqref{hol_Phi_eq} as the classical equation of motion, because
it is of first order in the time $\tau$. Instead, one should consider
\eqref{hol_Phi_eq} as the solution to the e.o.m, which in turn can
be deduced by $\tau$-differentiating \eqref{hol_Phi_eq}, 
\begin{equation}
\ddot{\Phi}-(\square+m^{2}+g\mathrm{U}'')(\square\Phi+m^{2}\Phi+g\mathrm{U}')=0.
\end{equation}
This e.o.m. can be derived from the action 
\begin{equation}
S=\intop_{0}^{+\infty}d\tau\int d^{4}x\frac{1}{2}\left(\dot{\Phi}^{2}+\left(\square\Phi+m^{2}\Phi+g\mathrm{U}'(\Phi)\right)^{2}\right),\label{hol_S}
\end{equation}
which indeed leads to $\dot{\Phi}$ as a canonically conjugate momentum
for $\Phi$. This $5d$ model is non-relativistic and contains higher-derivatives,
so its meaning is not immediately clear. But curiously, it mimics
some holographic features. Namely, if one evaluates the action \eqref{hol_S}
on its minimal trajectory \eqref{hol_Phi_eq}, then one gets $4d$
action of the underlying primary scalar $\phi(x)=\Phi(\tau=0,x)$
\begin{equation}
S^{on\textrm{-}shell}=\int d^{4}x\left(\frac{1}{2}\phi\square\phi+\frac{m^{2}}{2}\phi^{2}+\mathrm{U}(\phi)\right)+const,
\end{equation}
assuming that asymptotics $\Phi(\tau\rightarrow\infty)$ is fixed.
Thus, a classical $4d$ primary action arises as an on-shell $5d$
action \eqref{hol_S} treated as a functional of the initial value
$\phi$.

However, it is not straightforward to extend this relation to the
quantum level, because from the standpoint of $5d$ model the quantization
procedure we perform is far from the standard one: we quantize a classical
solution \eqref{hol_Phi_eq} by imposing equal-time commutation relation
\eqref{F_J_comm_int}, which is canonical only at the initial moment
$\tau=0$.

\section{Conclusion\label{SEC_CONCLUSIONS}}

In the paper we have studied the problem of quantization of the unfolded
system of the $4d$ scalar field with a self-interaction potential
of the general form. We have presented and analyzed corresponding
classical unfolded system, which is of interest in itself, since the
number of available nonlinear unfolded models is quite limited.

We have proposed three different but related ways of formulating unfolded
quantum field theory. All of them require classical off-shell unfolded
system as the starting point. The first one consists in imposing functional
Schwinger--Dyson equations as unfolded operator equations on the
partition function of the theory. This requires finding a consistent
commutation relation between operators of an unfolded field and an
unfolded source. This relation turns out to be quite remarkable: instead
of delta-function presented in a standard QFT, an unfolded commutator
represents (in the free case) a heat kernel, dependent on an auxiliary
variable $\tau$, which appears already in the classical unfolded
system, where it parameterizes off-shell descendants of the primary
scalar field. In the commutator of unfolded quantum fields, $\tau$
plays the role of a natural regularizer and thus one may hope that
unfolded dynamics approach will provide new instruments for dealing
with the problem of divergences in QFT. Another curious feature is
that the mentioned unfolded commutator becomes field-dependent in
the nonlinear theory, that reflects the nonlinearity of relations
between descendants and primaries in the unfolded module. We have
constructed this formulation and have used it to solve the free model,
while for nonlinear theories two other formulations seems more handy
and promising.

The second way to formulate unfolded QFT is in terms of the infinite
chain of unfolded Schwinger--Dyson equations for correlators, which
allows one to perturbatively calculate unfolded correlation functions.
However, to construct it, one needs an unfolded functional Schwinger--Dyson
system from the previous paragraph. We have presented a corresponding
unfolded correlators system and have used it to evaluate a first perturbative
correction to the unfolded propagator.

The third way to quantize unfolded field theory is to write down an
unfolded effective equations, which allows one to systematically restore
unfolded vertex functions within the framework of the semiclassical
expansion. Here the central object is the unfolded quantum Poisson
bracket. We have built a general unfolded effective system for an
arbitrary bracket and have analyzed a particular example, evaluating
a one-loop correction to the inverse propagator.

Finally, we have presented an auxiliary $5d$ model, which has $\tau$-variable
as the physical time, $\tau$-equation of the $4d$ unfolded system
as its classical solution and generates a correct $4d$ scalar action
as an on-shell $5d$ action evaluated as a functional of initial values
of the field. Although the status of this model is not entirely clear,
its very existence indicates that the auxiliary variable $\tau$ of
the unfolded dynamics approach may have some deeper meaning than a
cursory glance suggests.

In this paper, we have focused mainly on the problem of formulating
unfolded QFT, limiting ourselves to a few calculations for illustration
purposes. Therefore, it would be interesting to consider some concrete
scalar field theory in order to develop a systematic technique of
calculations in unfolded QFT, including Feynman diagrams, renormalizations
etc. In particular, it looks prominent to try to apply general heat-kernel
methods, considering that the heat kernel plays the central role in
formulating unfolded QFT.

On the other hand, the problem of formulating an unfolded quantum
gauge field theory raises new questions related to gauge symmetries,
ghosts etc. This is the next natural step towards the quantization
of the unfolded HS gravity, which is the ultimate goal of the presented
analysis. In this regard, it may be useful to relate the approach
of this paper to the approach of the Lagrange anchor \citep{Anchor3,Anchor2,Anchor4,Anchor1,Anchor5},
which is also aimed at non-Lagrangian quantization of field theories.
All this requires additional thorough analysis, which is beyond the
scope of this paper.

\section*{Acknowledgments}

The author is grateful to A.O. Barvinsky, V.E. Didenko, E.D. Skvortsov
and M.A. Vasiliev for valuable comments. The research was supported
by the Alexander von Humboldt Foundation.

\section*{Appendix A\label{Appendix A}. General unfolded frame for the scalar
field}

We consider a following general Ansatz for an unfolded equation of
a self-interacting scalar 
\begin{equation}
\mathrm{D}\Phi+a_{N}e\partial\bar{\partial}\Phi+b_{N}ey\bar{y}(\dot{\Phi}+m^{2}\Phi)+c_{N}ey\bar{y}g\mathrm{U}'(f_{N}\Phi)=0,
\end{equation}
where $a_{N}$, $b_{N}$, $c_{N}$, $f_{N}$ depend on Euler operator
$N$ \eqref{Euler} and $\mathrm{U}'$ is understood as a formal series
\begin{equation}
\mathrm{U}'(f_{N}\Phi):=\sum_{n=2}^{\infty}\frac{u_{n}}{n!}(f_{N}\Phi)^{n},
\end{equation}
with every $f_{N}$ acting only on the one following $\Phi$.

An unfolded consistency condition, arising from $\mathrm{D}^{2}\equiv0$,
requires 
\begin{equation}
b_{N}=\frac{b}{N(N+1)a_{N-1}},
\end{equation}
\begin{equation}
c_{N}=\frac{c}{(N+1)!(a_{0}\cdot a_{1}\cdot...\cdot a_{N-1})},
\end{equation}
\begin{equation}
f_{N}=f\cdot N!(a_{0}\cdot a_{1}\cdot...\cdot a_{N-1}).
\end{equation}
Then $Y$-dependence of $\Phi$ is resolved as 
\begin{equation}
\Phi(Y,\tau|x)=\sum_{n=0}^{\infty}\frac{(-y^{\alpha}\bar{y}^{\dot{\alpha}}\nabla_{\alpha\dot{\alpha}})^{n}}{(n!)^{2}(a_{0}\cdot...\cdot a_{n-1})}\Phi(0,\tau|x),
\end{equation}
and $\tau$-dependence is determined by 
\begin{equation}
\square\Phi-m^{2}\Phi-c_{N}(N+1)\mathrm{U}'(f_{N}\Phi)=\dot{\Phi}.
\end{equation}

In the paper we pick up a particular solution 
\begin{equation}
a_{N}=\frac{1}{N+1},\:b_{N}=\frac{1}{N+1},\:c_{N}=\frac{1}{N+1},\:f_{N}=1,\label{int_choice}
\end{equation}
but different choices are also possible. In the HS literature a standard
choice for unfolded equations is to demand $a_{N}=1$. For the model
under consideration this choice entails (up to overall scaling of
variables) 
\begin{equation}
a_{N}=1,\:b_{N}=\frac{1}{N(N+1)},\:c_{N}=\frac{1}{(N+1)!},\:f_{N}=N!,\label{stan_choice}
\end{equation}
so that $Y$-dependence is resolved as 
\begin{equation}
\Phi(Y,\tau|x)=\text{}_{0}F_{1}(;1;y^{\alpha}\bar{y}^{\dot{\alpha}}\nabla_{\alpha\dot{\alpha}})\Phi(0,\tau|x),
\end{equation}
where $\text{}_{0}F_{1}(;1;z)$ is a confluent hypergeometric limit
function, which can also be expressed through the modified Bessel
function as $\text{}_{0}F_{1}(;1;z)=I_{0}(2\sqrt{z})$, and $\tau$-equation
takes the form 
\begin{equation}
\square\Phi-m^{2}\Phi-\frac{1}{N!}g\mathrm{U}'(N!\Phi)=\dot{\Phi}.
\end{equation}
We see that the solution \eqref{int_choice} has two important advantages
over \eqref{stan_choice}: first, $Y$-dependence comes down to a
simple shift of $x$-coordinate by $y\bar{y}$ as seen from \eqref{Y_shift};
second, $Y$- and $\tau$-dependencies are completely separated and
unfolded potential arises by trivially replacing the primary field
$\phi$ with the unfolded field $\Phi$.


\begin{thebibliography}{99}
\bibitem{NoGo} X. Bekaert, N. Boulanger, P. Sundell, \emph{Rev.Mod.Phys.}
\textbf{84} (2012) 987-1009 \href{https://arxiv.org/abs/1007.0435}{[arXiv:1007.0435]}.

\bibitem{FradVas}E.S. Fradkin, M.A. Vasiliev, \emph{Annals Phys.}
\textbf{177} (1987) 63.

\bibitem{snow}X. Bekaert, N. Boulanger, A. Campoleoni, M. Chiodaroli,
D. Francia, M. Grigoriev, E. Sezgin, E. Skvortsov, \emph{Snowmass
White Paper: Higher Spin Gravity and Higher Spin Symmetry}, \href{https://arxiv.org/abs/2205.01567}{[arXiv:2205.01567]}.

\bibitem{unf1}M.A. Vasiliev, \emph{Annals Phys.} \textbf{190} (1989)
59-106.

\bibitem{vas1}M.A. Vasiliev, \emph{Phys.Lett.B} \textbf{243} (1990)
378-382.

\bibitem{vas2}M.A. Vasiliev, \emph{Phys.Lett.B} \textbf{285} (1992)
225-234.

\bibitem{unf2}M.A. Vasiliev, \emph{Class.Quant.Grav.} \textbf{11}
(1994) 649-664.

\bibitem{ActionsCharges}M.A. Vasiliev, \emph{Int.J.Geom.Meth.Mod.Phys.}
\textbf{3} (2006) 37-80 \href{https://arxiv.org/abs/hep-th/0504090}{[hep-th/0504090]}.

\bibitem{hs_3d}S.F. Prokushkin, M.A. Vasiliev, \emph{Nucl.Phys.B}
\textbf{545} (1999) 385 \href{https://arxiv.org/abs/hep-th/9806236}{[hep-th/9806236]}.

\bibitem{hs_arb_d}M.A. Vasiliev, \emph{Phys.Lett.B} \textbf{567}
(2003) 139-151 \href{https://arxiv.org/abs/hep-th/0304049}{[hep-th/0304049]}.

\bibitem{xir1}A. Sharapov, E. Skvortsov, A. Sukhanov, R. Van Dongen,
\emph{JHEP }\textbf{09} (2022) 134, \emph{JHEP} \textbf{02} (2023)
183 (erratum)\emph{ }\href{https://arxiv.org/abs/2205.07794}{[arXiv:2205.07794]}.

\bibitem{xir2}A. Sharapov, E. Skvortsov, \emph{Nucl.Phys.B}\textbf{
985 }(2022) 115982\emph{ }\href{https://arxiv.org/abs/2205.15293}{[arXiv:2205.15293]}.

\bibitem{act1}S.R. Das, A. Jevicki, \emph{Phys.Rev.D} \textbf{68}
(2003) 044011 \href{https://arxiv.org/abs/hep-th/0304093}{[hep-th/0304093]}.

\bibitem{act2}A. Fotopoulos, M. Tsulaia, \emph{Int.J.Mod.Phys.A}
\textbf{24} (2009) 1-60 \href{https://arxiv.org/abs/0805.1346}{[arXiv:0805.1346]}.

\bibitem{act3}A. Jevicki, K. Jin, Q. Ye, \emph{J.Phys.A} \textbf{44}
(2011) 465402 \href{https://arxiv.org/abs/1106.3983}{[arXiv:1106.3983]}.

\bibitem{act4}N. Boulanger, P. Sundell, \emph{J.Phys.A} \textbf{44}
(2011) 495402 \href{https://arxiv.org/abs/1102.2219}{[arXiv:1102.2219]}.

\bibitem{act5}N. Boulanger, N. Colombo, P. Sundell, \emph{JHEP} \textbf{10}
(2012) 043 \href{https://arxiv.org/abs/1205.3339}{[arXiv:1205.3339]}.

\bibitem{act6}N. Boulanger, E. Sezgin, P. Sundell, \emph{4D Higher
Spin Gravity with Dynamical Two-Form as a Frobenius-Chern-Simons Gauge
Theory} \href{https://arxiv.org/abs/1505.04957}{[arXiv:1505.04957]}.

\bibitem{act7}I.L. Buchbinder, K. Koutrolikos, \emph{JHEP} \textbf{12}
(2015) 106 \href{https://arxiv.org/abs/1510.06569}{[arXiv:1510.06569]}.

\bibitem{act8}C. Arias, R. Bonezzi, N. Boulanger, E. Sezgin, P. Sundell,
in \emph{Proccedings of International Workshop on Higher Spin Gauge
Theories, 4-6 November 2015, Singapore} (2017) 213-253, \href{https://arxiv.org/abs/1603.04454}{[arXiv:1603.04454]}.

\bibitem{Z1}S. Giombi, I.R. Klebanov, \emph{JHEP} \textbf{12} (2013)
068 \href{https://arxiv.org/abs/1308.2337}{[arXiv:1308.2337]}.

\bibitem{Z2}S. Giombi, I.R. Klebanov, B.R. Safdi, \emph{Phys.Rev.D}
\textbf{89} (2014) 8, 084004 \href{https://arxiv.org/abs/1401.0825}{[arXiv:1401.0825]}.

\bibitem{Z4}S. Giombi, I.R. Klebanov, A.A. Tseytlin, \emph{Phys.Rev.D}
\textbf{90} (2014) 2, 024048 \href{https://arxiv.org/abs/1402.5396}{[arXiv:1402.5396]}.

\bibitem{Z5}M. Beccaria, A.A. Tseytlin, \emph{JHEP} \textbf{11} (2014)
114 \href{https://arxiv.org/abs/1410.3273}{[arXiv:1410.3273]}.

\bibitem{Z6}M. Beccaria, A.A. Tseytlin, \emph{J.Phys.A} \textbf{48}
(2015) 27, 275401 \href{https://arxiv.org/abs/1503.08143}{[arXiv:1503.08143]}.

\bibitem{Z7}M. Beccaria, A. A. Tseytlin, \emph{J.Phys.A} \textbf{49}
(2016) 29, 295401, \href{https://arxiv.org/abs/1602.00948}{[arXiv:1602.00948]}.

\bibitem{Z8}Yi Pang, E. Sezgin, Y. Zhu, \emph{Phys.Rev.D} \textbf{95}
(2017) 2, 026008 \href{https://arxiv.org/abs/1608.07298}{[arXiv:1608.07298]}.

\bibitem{Z9}S. Giombi, I.R. Klebanov, Z.M. Tan, \emph{Universe} \textbf{4}
(2018) 1, 18 \href{https://arxiv.org/abs/1608.07611}{[arXiv:1608.07611]}.

\bibitem{Z10}E.D. Skvortsov, T. Tran, \emph{Universe} \textbf{3}
(2017) 3, 61 \href{https://arxiv.org/abs/1707.00758}{[arXiv:1707.00758]}.

\bibitem{amp0}D. Ponomarev, A.A. Tseytlin, \emph{JHEP} \textbf{05}
(2016) 184, \href{https://arxiv.org/abs/1603.06273}{[arXiv:1603.06273]}.

\bibitem{amp1}S. Giombi, C. Sleight, M. Taronna, \emph{JHEP} \textbf{06}
(2018) 030 \href{https://arxiv.org/abs/1708.08404}{[arXiv:1708.08404]}.

\bibitem{amp2}C. Sleight, M. Taronna, \emph{JHEP} \textbf{01} (2018)
060 \href{https://arxiv.org/abs/1708.08668}{[arXiv:1708.08668]}.

\bibitem{amp3}D. Ponomarev, E. Sezgin, E. Skvortsov, \emph{JHEP}
\textbf{11} (2019) 138 \href{https://arxiv.org/abs/1904.01042}{[arXiv:1904.01042]}.

\bibitem{amp4}B. Nagaraj, D. Ponomarev, \emph{Phys.Rev.Lett.} \textbf{122}
(2019) 10, 101602 \href{https://arxiv.org/abs/1811.08438}{[arXiv:1811.08438]}.

\bibitem{amp6}B. Nagaraj, D. Ponomarev, \emph{JHEP} \textbf{06} (2020)
068 \href{https://arxiv.org/abs/1912.07494}{[arXiv:1912.07494]}.

\bibitem{amp7}B. Nagaraj, D. Ponomarev, \emph{JHEP} \textbf{08} (2020)
08, 012 \href{https://arxiv.org/abs/2004.07989}{[arXiv:2004.07989]}.

\bibitem{chir1}D. Ponomarev, E.D. Skvortsov, \emph{J.Phys.A} \textbf{50}
(2017) 9, 095401 \href{https://arxiv.org/abs/1609.04655}{[arXiv:1609.04655]}.

\bibitem{chir2}D. Ponomarev, \emph{JHEP} \textbf{12} (2016) 117 \href{https://arxiv.org/abs/1611.00361}{[arXiv:1611.00361]}.

\bibitem{chir3}D. Ponomarev, \emph{JHEP} \textbf{12} (2017) 141 \href{https://arxiv.org/abs/1710.00270}{[arXiv:1710.00270]}.

\bibitem{chir4}E.D. Skvortsov, T. Tran, M. Tsulaia, \emph{Phys.Rev.Lett.}
\textbf{121} (2018) 3, 031601 \href{https://arxiv.org/abs/1805.00048}{[arXiv:1805.00048]}.

\bibitem{chir5}E. Skvortsov, T. Tran, \emph{JHEP} \textbf{07} (2020)
021 \href{https://arxiv.org/abs/2004.10797}{[arXiv:2004.10797]}.

\bibitem{chir6}E. Skvortsov, T. Tran, M. Tsulaia, \emph{Phys.Rev.D}
\textbf{101} (2020) 10, 106001 \href{https://arxiv.org/abs/2002.08487}{[arXiv:2002.08487]}.

\bibitem{misuna}N.G. Misuna, \emph{Phys.Lett.B} \textbf{798} (2019)
134956 \href{https://arxiv.org/abs/1905.06925}{[arXiv:1905.06925]}.

\bibitem{misuna-1}N.G. Misuna, \emph{JHEP} \textbf{12} (2021) 172
\href{https://arxiv.org/abs/2012.06570}{[arXiv:2012.06570]}.

\bibitem{misuna3}N.G. Misuna, \emph{Phys.Lett.B }\textbf{840} (2023)
137845 \href{https://arxiv.org/abs/2201.01674}{[arXiv:2201.01674]}.

\bibitem{Anchor1}D.S. Kaparulin, S.L. Lyakhovich, A.A. Sharapov,
\emph{Int.J.Mod.Phys.A} \textbf{26} (2011) 1347-1362 \href{https://arxiv.org/abs/1012.2567}{[arXiv:1012.2567]}.

\bibitem{Vas_StarProduct}M. A. Vasiliev, \emph{Higher spin gauge
theories: Star product and AdS space}, In {*}Shifman, M.A. (ed.):
The many faces of the superworld{*} 533-610 \href{https://arxiv.org/abs/hep-th/9910096}{[hep-th/9910096]}.

\bibitem{elem_vas}V.E. Didenko, E.D. Skvortsov, \emph{Elements of
Vasiliev theory} \href{https://arxiv.org/abs/1401.2975}{[arXiv:1401.2975]}.

\bibitem{Anchor2}S.L. Lyakhovich, A.A. Sharapov, \emph{JHEP} \textbf{02}
(2006) 007 \href{https://arxiv.org/abs/hep-th/0512119}{[hep-th/0512119]}.

\bibitem{unf_Lagr}A.A. Tarusov, M.A. Vasiliev, \emph{Phys.Lett.B}
\textbf{825} (2022) 136882 \href{https://arxiv.org/abs/2111.12691}{[arXiv:2111.12691]}.

\bibitem{Lee_q_off_shell}K. Lee, \emph{JHEP} \textbf{05} (2022) 05
\href{https://arxiv.org/abs/2202.08133}{[arXiv:2202.08133]}.

\bibitem{Anchor3}P.O. Kazinski, S.L. Lyakhovich, A.A. Sharapov \emph{JHEP}
\textbf{07} (2005) 076 \href{https://arxiv.org/abs/hep-th/0506093}{[hep-th/0506093]}.

\bibitem{Anchor4} S.L. Lyakhovich, A.A. Sharapov, \emph{JHEP} \textbf{01}
(2007) 047 \href{https://arxiv.org/abs/hep-th/0612086}{[hep-th/0612086]}.

\bibitem{Anchor5} D.S. Kaparulin, S.L. Lyakhovich, A.A. Sharapov,
\emph{SIGMA} \textbf{8} (2012) 021 \href{https://arxiv.org/abs/1112.1860}{[1112.1860]}.

\end{thebibliography}
\end{document}